\documentclass[11pt]{article}
\usepackage{amsfonts}
\usepackage{dsfont}
\usepackage{physics}
\usepackage{amsmath,amssymb}
\usepackage{bbm}
\usepackage{bm}
\usepackage{color}
\usepackage{slashed}
\usepackage{cite}
\usepackage[utf8]{inputenc}
\usepackage{graphicx}
\usepackage{bbold}
\DeclareGraphicsExtensions{.pdf,.png,.jpg}

\newcommand*{\cf}{cf.\ }
\newcommand*{\ie}{i.e.\ }
\newcommand*{\eg}{e.g.\ }
\newcommand*{\Eq}{eq.\,}
\newcommand*{\Eqs}{eqs.\,}

\usepackage[colorlinks=false,citecolor=cyan,urlcolor=blue,bookmarks=true,bookmarks=true,bookmarksopen=true,bookmarksnumbered=true,bookmarksopenlevel=3]{hyperref}

\definecolor{airforceblue}{rgb}{0.36, 0.54, 0.66}
\definecolor{steelblue}{rgb}{0.27, 0.51, 0.71}
\definecolor{amber}{rgb}{1.0, 0.49, 0.0}

\def\simg{{\ \lower-1.2pt\vbox{\hbox{\rlap{$>$}\lower6pt\vbox{\hbox{$\sim$}}}}\ }}
\def\siml{{\ \lower-1.2pt\vbox{\hbox{\rlap{$<$}\lower6pt\vbox{\hbox{$\sim$}}}}\ }}

\makeatletter \@addtoreset{equation}{section} \makeatother

\def\simg{{\ \lower-1.2pt\vbox{\hbox{\rlap{$>$}\lower6pt\vbox{\hbox{$\sim$}}}}\ }}
\def\siml{{\ \lower-1.2pt\vbox{\hbox{\rlap{$<$}\lower6pt\vbox{\hbox{$\sim$}}}}\ }}

\makeatletter \@addtoreset{equation}{section} \makeatother

\begin{document}

\flushbottom

\begin{titlepage}

\begin{centering}

\vfill

\begin{flushright}
TTP21-018, P3H-21-042
\end{flushright}

{\Large{\bf
Non-relativistic and potential non-relativistic 
\\
effective field theories  for scalar mediators
}} 

\vspace{0.8cm}

S.~Biondini$^a$ and V.~Shtabovenko$^b$

\vspace{0.8cm}

$^{a}${\em Department of Physics, University of Basel,
\\
 Klingelbergstr. 82, CH-4056 Basel, Switzerland} 
\\
\vspace{0.15 cm}
$^{b}${\em Institut für Theoretische Teilchenphysik (TTP),\\
Karlsruhe Institute of Technology (KIT), 
\\
Wolfgang-Gaede-Straße 1, 76131 Karlsruhe, Germany}
\\
\vspace*{0.8cm}

\end{centering}

\vspace*{0.3cm}
 
\noindent

\textbf{Abstract}: Yukawa-type interactions between heavy Dirac fermions and a scalar field are a common 
	ingredient in various extensions of the Standard Model. Despite of that, the non-relativistic
	limit of the scalar Yukawa theory has not yet been studied in full generality in a rigorous 
	and model-independent way. In this paper we intend to fill this gap by initiating a series of
	investigations that make use of modern effective field theory (EFT) techniques. In particular, 
	we aim at constructing suitable non-relativistic and potential non-relativistic EFTs of Yukawa
	interactions (denoted as NRY and pNRY respectively) in close analogy to the well known and 
	phenomenologically successful non-relativistic QCD (NRQCD) and potential non-relativistic QCD
	(pNRQCD). The phenomenological motivation for our study lies in the possibility to explain
	the existing cosmological observations by introducing heavy fermionic dark matter particles
	that interact with each other by exchanging a light scalar mediator. A systematic study of
	this compelling scenario in the framework of non-relativistic EFTs (NREFTs) constitutes the 
	main novelty of our approach as compared to the existing studies.


\vfill
\newpage
\tableofcontents
\end{titlepage}

\section{Introduction}
Dark sectors containing light vectors or scalars may feature sizable self-interactions between dark matter (DM) particles and are therefore of high phenomenological interest. Self-interacting dark matter appears to reproduce the observed galactic structure better than collisionless DM \cite{Spergel:1999mh,Kusenko:2001vu,Feng:2008mu,Loeb:2010gj,Weinberg:2013aya,Peter:2012jh,Rocha:2012jg} and may offer a dynamical explanation for the scaling relations governing galactic halos all the way up to clusters of galaxies \cite{Foot:2013nea,Foot:2013lxa,Foot:2013uxa,Markevitch:2003at,Randall:2007ph,Kahlhoefer:2013dca,Harvey:2015hha,Kaplinghat:2015aga}. A simultaneous description of small scale objects as well as large-scale formations (\eg dwarf galaxies and galaxy clusters respectively) requires velocity-dependent DM self-interactions that are naturally achieved in models featuring a light DM mediator \cite{Buckley:2009in,Feng:2009hw,Feng:2009mn,Loeb:2010gj,Kaplinghat:2015aga,Ackerman:mha,Aarssen:2012fx}. 

On top of being desirable from the phenomenological and observational points of views, the possibility of a richer dark sector, that comprises more than one particle, is fairly common in many DM models, \cf \eg \cite{Bertone:2004pz,DeSimone:2016fbz,Petraki:2013wwa}. The dark particles can enjoy their own hidden forces, which are far less constrained than the interactions between DM and Standard Model (SM) degrees of freedom.
Furthermore, the existence of light (\ie with masses much smaller than that of the actual DM particles) mediators may affect the DM dynamics in multiple ways. Most notably, whenever DM particles are slowly moving with non-relativistic velocities, light mediators can induce bound states in the dark sector in the  early universe and/or  in  the  dense  environment  of  present-day haloes \cite{Hisano:2002fk,Hisano:2003ec,Feng:2009mn,vonHarling:2014kha}. As for the above-threshold states, the effect of repeated mediator exchange manifests itself in the so-called Sommerfeld enhancement for an attractive potential \cite{Sommerfeld:1931,Sakharov:1991pia}. In this context the role of a light mediator can also be played by SM particles. For a sufficiently heavy DM these may even be weak gauge bosons \cite{Hisano:2002fk,Hisano:2003ec,Cirelli:2007xd,Beneke:2014gja,Beneke:2016ync,Mitridate:2017izz} or the Higgs boson \cite{Harz:2017dlj,Beneke:2014gja}. This latter option is becoming increasingly relevant as null searches for new physics at the LHC are pushing the scale of possible novel particles, including many thermally produced DM candidates, into the multi-TeV region.\footnote{The complementary alternative is to consider light and ultra-light DM, \cf \eg \cite{Duffy:2009ig,Ferreira:2020fam,Niemeyer:2019aqm}.}

Depending on the model at hand, one may find unstable bound states, that usually appear in symmetric DM models, as well as stable bound states (the latter are part of the present-day DM energy density). Typically, the annihilating particle-antiparticle pairs feel an attractive potential that can not only drastically change the annihilation cross section via Sommerfeld enhancement
but also induce bound-state formation \cite{Feng:2009mn,vonHarling:2014kha}. Once bound states are formed, and not effectively dissociated in the thermal plasma, they provide an additional channel for the depletion of DM particles in the early universe. The relic density determination has to be adjusted accordingly, since substantial annihilations may still occur after the chemical decoupling. This typically results in (i) mapping out different combinations of DM masses and couplings that reproduce the observed DM cosmological abundance $\Omega_{\hbox{\tiny DM}}h^2=0.1200 \pm 0.0012$ \cite{Aghanim:2018eyx}; (ii) a reinvestigation of DM phenomenology due to the interplay between the model parameters that fix the relic density and guide the experimental strategies. The stable bound states that often arise in asymmetric DM models affect the detection strategy and experimental searches for both indirect \cite{MarchRussell:2008tu,Pearce:2013ola,Pearce:2015zca,Frandsen:2014lfa,Boddy:2014qxa} and direct detection signals \cite{Laha:2013gva}. 

The impact of bound-state effects on the DM relic density is, of course, model dependent. When accounting for both Sommerfeld enhancement and bound-state formation, the DM mass compatible with the observed energy density can change from a few per-cent level up to one order of magnitude. Much of the recent literature is focused on vector mediators, for which a comprehensive  and diversified refinements for deriving bound-state formation cross sections at zero \cite{Feng:2009mn,vonHarling:2014kha,Beneke:2016ync,Liew:2016hqo,Cirelli:2007xd,Mitridate:2017izz,Harz:2019rro} and  finite temperature \cite{Kim:2016kxt,Biondini:2017ufr,Biondini:2018pwp,Biondini:2018ovz,Binder:2018znk,Biondini:2019int,Binder:2019erp,Biondini:2019zdo} have been carried out together with the impact on the relic density, and on model phenomenology/experimental prospects \cite{Biondini:2018ovz,Biondini:2019int,Bottaro:2021srh}. Only recently a more systematic study of the role and impact of scalar mediators with respect to bound-state formation has been initiated \cite{Wise:2014jva,Petraki:2015hla,An:2016kie,Biondini:2018xor,Oncala:2018bvl,Harz:2019rro,Oncala:2019yvj}.

In this work we adopt an effective field theory (EFT) approach to address the bound-state dynamics of heavy DM particles. Indeed, the problem at hand comes as a multi-scale system. On the one hand, one finds three typical scales of a non-relativistic dynamics, which are assumed to be well separated, namely $M \gg Mv \gg Mv^2$, where $M$ is the DM particle mass, while $v$ denotes its typical velocity in a bound state. A Coulombic bound state satisfies $v \sim \alpha$, with $\alpha$ being the relevant coupling constant. In addition to these scales our system also contains the mediator mass $m$, which we assume to be much lighter than the DM mass, and thermal scales, most notably the temperature of the early universe plasma and thermal masses. In particular, we shall employ the framework of non-relativistic effective field theories \cite{Caswell:1985ui,Bodwin:1994jh} (NREFTs) and potential non-relativistic effective field theories \cite{Pineda:1997bj,Brambilla:1999xf} (pNREFTs), which are obtained by \emph{integrating out} energy/momenta of order $M$ and $Mv$ respectively. In doing so we can construct suitable low-energy EFTs describing the degrees of freedom we are interested in. These are DM fermion pairs, either in bound or scattering states, and low-energetic scalar mediators. Bound-state calculations can be then carried out in a very similar way to the ordinary quantum mechanics, with the important difference that higher order corrections to the potentials, and other observables, can be obtained in a systematic and model-independent way from quantum field theoretical \emph{matching} calculations. Our approach is based on the renowned NREFTs of this sort that have been obtained for QED and QCD, and served as precious and handy tools for rigorous and systematic analyses of \eg hydrogen atom, positronium, heavy quarkonia, heavy-light hadrons or muonic hydrogen (we refer to \cite{Brambilla:2004jw,Brambilla:2010cs,Brambilla:2014jmp} for an overview of the existing results in the context of strong interactions).

A non-relativistic scalar Yukawa theory (NRY) constructed in the spirit of \cite{Caswell:1985ui,Bodwin:1994jh} has been already  considered in \cite{Luke:1996hj,Luke:1997ys}. There it was essentially employed as a toy-model to illustrate some concepts of the NRQCD power-counting and the rationale of applying NREFTs to bound states. A systematic study of the nonrelativistic dark matter in the framework of the minimal supersymmetric standard model (MSSM) using NREFT techniques was carried out in \cite{Beneke:2012tg,Hellmann:2013jxa,Beneke:2014gja}. We also would like to point out that the fermion-bilinear sector of the \emph{pseudoscalar} Yukawa theory at $\mathcal{O}(1/M^3)$ can be found in \cite{Platzman:1960dqa}. Of course, since \cite{Platzman:1960dqa} was published long before the EFT techniques became mainstream, the derivation presented there does not use the modern language and methods of NREFTs. Apart from the NRY we also consider the pNREFT version of the scalar Yukawa theory, which we call potential non-relativistic scalar Yukawa theory (pNRY). It is worth noting that the effect of adding interactions between heavy fermions and the Higgs to the conventional pNRQCD (which naturally leads to Yukawa potentials) has been considered e.\,g. in \cite{Eiras:2006xm,Beneke:2015lwa} when studying $t\bar{t}$-production near threshold. 

At variance with the previous works, here we are interested in the pure Yukawa theory that lacks any interactions with gauge bosons such as photons, gluons, $W$ or $Z$. Furthermore, we would like to abstain from introducing any additional symmetries apart from what is already present in the scalar Yukawa theory. In our view, this approach allows us to investigate and highlight the essential features of non-relativistic scalar Yukawa interactions in a 
clear and transparent fashion without making any assumptions on the nature of the underlying higher-energy theory. The aim of the present work is to revisit the construction of the NRY by extending the treatment of \cite{Luke:1996hj,Luke:1997ys} and to explore the consequences of the resulting NREFT and pNREFT for the DM phenomenology, where
we are interested in describing the interactions of heavy Dirac fermions $X$ with a much lighter scalar field $\phi$. 
To the best of our knowledge, pNRY as a pNRQED-like theory that contains solely Yukawa interactions is presented in this work for the first time. In both cases we explore possible hierarchies of scales and discuss the appropriate power-counting rules. In this paper, we shall focus on the zero temperature case, and only marginally comment on the finite temperature generalization. 

It is worth noting that the DM model under consideration has some intriguing properties that are unique to heavy fermions exchanging a scalar. First, as opposed to the vector mediator case, the annihilation of heavy particle-antiparticle pairs
at leading order in the velocity and $1/M$ expansion proceeds via a $P$-wave process. More explicitly, one finds that the matching coefficients of 4-fermion dimension-6 operators vanish at $\mathcal{O}(\alpha^2 v^0)$, whereas the first non-vanishing contributions show up in the velocity suppressed dimension-8 operators. Second, the pNRY exhibits, already at the Lagrangian level, the absence of electric-dipole transitions and the presence of \emph{monopole} and \emph{quadrupole} interactions between a heavy pair and the  scalar mediator. In the context of pNREFTs, monopole interactions were discussed for super-symmetric Yang-Mills theories at weak coupling in \cite{Pineda:2007kz}. Finally, in the case of vector mediators, pNREFTs have been already fully, or at least to some extent, exploited in the context of DM with and without co-annihilating partners  \cite{Asadi:2016ybp,Biondini:2018ovz,Biondini:2019int,Beneke:2020vff,Binder:2019erp,Binder:2020efn}. 

The structure of the paper is as follows. In section \ref{sec_model} we briefly introduce the simplified model that we take as our high-energy (in the EFT sense) theory. Then, in section \ref{sec:NREFTs_general} we address the construction of the low-energy NREFT (denoted as NRY) for non-relativistic fermions and antifermions exchanging a scalar. Here we shall give the set of operators as an expansion in $1/M$, $v$ and coupling constants, and discuss the symmetries and power counting rules of the low-energy theory. In section \ref{sec_NREFT} we apply the NRY formalism to describe DM interactions and provide the results for the matching coefficients. As far as the fermion bilinears are concerned, we shall be content with tree-level matching coefficients. The matching for 4-fermion dimension-6 and dimension-8 operators will be carried out at $\mathcal{O}(\alpha^2)$. These operators encode the hard contribution to the annihilation cross section for the process $X \bar{X} \to \phi \phi$. In section \ref{sec_pNREFT} we proceed to the derivation of the pNREFT (denoted as pNRY), whose degrees of freedom are bound states, scattering states with kinetic energy of order $M v^2$ and ultrasoft scalar particles. We perform the potential matching at $\mathcal{O}(M \alpha^4)$ and then provide an application of pNRY to the derivation of the discrete spectrum and the calculation of the bound-state formation cross section. Conclusions and outlook are offered in section \ref{sec_conclusions}. 

\section{Dark matter model}
\label{sec_model}
In this section we briefly introduce the DM model under consideration and discuss the relevant degrees of freedom. We assume DM to be a Dirac fermion singlet under the SM gauge group that it is coupled to a scalar particle with a Yukawa-type interaction. The Lagrangian density of the model reads \cite{Pospelov:2007mp,Kaplinghat:2013yxa} 
\begin{equation}
    \mathcal{L}= \bar{X} (i \slashed{\partial} -M)X + \frac{1}{2} \partial_\mu \phi \, \partial^\mu \phi -\frac{1}{2}m^2 \phi^2 - \frac{\lambda}{4!} \phi^4 - g \bar{X} X \phi +\mathcal{L}_{\hbox{\scriptsize portal}} \, ,
    \label{lag_mod_0}
    \end{equation}
where $X$ is the DM Dirac field and $\phi$ is a real scalar field. The scalar self-coupling and the Yukawa coupling between the fermion and the scalar fields are denoted as $\lambda$ and $g$ respectively. The mass of the scalar mediator $m$ is assumed to be much smaller than the DM particle mass $M$, $m \ll M$. Here we adopt a simplified model realization, where the question of the fermion mass generation and of the gauge group governing the dark sector are ignored.\footnote{One can find a detailed and comprehensive study for a simplified model with two mediators, scalar and vector, in \cite{Kahlhoefer:2015bea,Duerr:2016tmh}, where the gauge invariance and spontaneous symmetry breaking in the dark sector is fully accounted for.} Our aim is to consider the Lagrangian given in \Eq\eqref{lag_mod_0} as one of the simplest representatives for a family of minimal DM models \cite{Kaplinghat:2013yxa,DeSimone:2016fbz} with a light scalar mediating interactions between DM particles. It goes without saying that such a scenario admits different realizations that can be much more involved than a single Yukawa interaction (\cf \eg \cite{Wise:2014jva,Kahlhoefer:2017umn,Oncala:2018bvl}). 

Next, $\mathcal{L}_{\hbox{\scriptsize portal}}$ accounts for the interactions between the scalar $\phi$ and other degrees of freedom that can be either in the dark sector (\eg all particles lighter than $\phi$), and/or in the SM sector. The most common realization of such a portal involves interactions with the SM Higgs boson. In general, portal interactions are needed because the light scalar particles $\phi$ are abundant in the early universe and a substantial population is still present after the freeze-out of the dark fermion. Hence, there has to be a mechanism that allows $\phi$ particles to decay and deplete their population  so
that the scalar does not happen to dominate the energy density of the Universe \cite{Kaplinghat:2013yxa,DelNobile:2015uua}.
The minimal model of \Eq\eqref{lag_mod_0} is moderately under tension if one considers the interactions of the scalar $\phi$ with the Higgs boson and hence SM fermions. Especially the interactions with quarks severely constrain the model via direct detection experiments \cite{Kaplinghat:2013yxa,Kainulainen:2015sva}. However, these tensions can be removed in a number of different ways \cite{Kahlhoefer:2017umn,Hambye:2019tjt}. Since a detailed phenomenological analysis is beyond the scope of this  work, we do not specify  $\mathcal{L}_{\hbox{\scriptsize portal}}$ further and merely focus on the complementary terms in \Eq\eqref{lag_mod_0} to derive the low-energy field theories relevant for the bound-state dynamics. This sets the stage for our NREFT and pNREFT formulations and paves the way for more thorough investigations (also with respect to the DM phenomenology) in future works.

\section{Non-relativistic Yukawa theory}
\label{sec:NREFTs_general}
In the following we would like to discuss the procedure of constructing a tower of non-relativistic EFTs for a heavy Dirac fermion $X$ that interacts with a light scalar field $\phi$ via a scalar Yukawa interaction. Our main motivation is to investigate the properties of $X \bar{X}$ bound states such as spectra, production and decays in a rigorous and model-independent way. In order to proceed systematically, it is useful to disentangle low-energy modes relevant for the bound-state formation from high-energy modes that are naturally present in the UV-complete theory described by \Eq\eqref{lag_mod_0}. Nevertheless, the contributions from large energies and momenta are not simply discarded: their effects will be incorporated into Wilson coefficients multiplying the operators that appear in the EFT Lagrangians. The process of determining these coefficients by comparing Green's functions of two theories at low-energies is called \emph{matching}.
The EFT description can be systematically improved by including higher-order operators compatible with the symmetries of the underlying theory. The effects of these operators can be quantified using EFT \emph{power-counting rules}, so that at each order in the relevant expansion parameters only a finite number of operators must be taken into account. This leads to a comprehensive description of the low-energy physics, that allows us to make predictions for the physical observables of interest (\eg cross sections or decay rates) in a simple and straightforward fashion.
\begin{figure}[t!]
    \centering
    \includegraphics[scale=0.55]{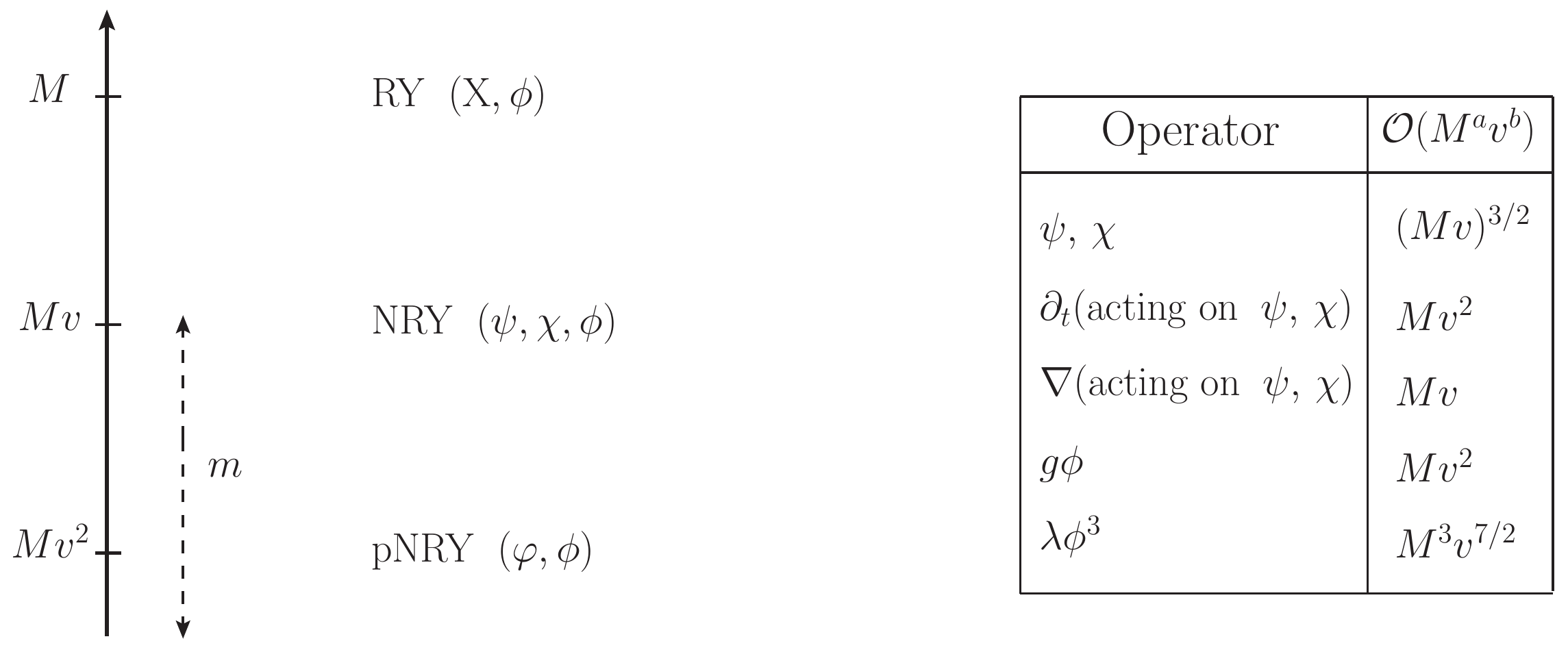}
    \caption{(Left) Hierarchy of scales for a non-relativistic Yukawa theory with $M \gg Mv \gg Mv^2$. The mass of the scalar mediator is assumed to be much smaller than the DM mass. The EFTs (together with their field content) that arise from integrating out the scales $M$ and $Mv$ are NRY and pNRY respectively. (Right) Estimate of the operator scaling in NRY.}
    \label{fig:NRYT_scales}
\end{figure}

Obviously, we need to assume a certain hierarchy between the scales relevant for the non-relativistic bound states (see figure \ref{fig:NRYT_scales}). The largest of these relevant scales is the heavy fermion mass  $M$. An important scale below $M$ is the typical size of the relative momentum between the fermions in a bound state, $|\bm{p}| \sim M v$, where $v$ is the relative velocity of the particles. Notice that this scale is also related to the typical bound state size $r$, where $1/r \sim M v$ (one can use the Bohr radius $a_0$ for Coulombic bound states for the size estimate). As our fermions are heavy and non-relativistic, we have $v \ll c$. We assume that $v$ is sufficiently small with at least $v^2 \leq 0.3$. In nature $v^2 \sim 0.3$ is found \eg in heavy quarkonia made of a charm and an anti-charm quarks. The non-relativistic description is still applicable to such systems, but the velocity expansion converges rather slowly. On the other hand, for $X \bar{X}$ bound states with  $v^2 \sim 0.1$ (as in $b \bar{b}$-quarkonia), corrections of $\mathcal{O}(v^2)$
should be sufficient for a reliable phenomenological analysis. The typical bound state energy of an $X \bar{X}$ system scales as $M v^2$. In the following we denote the scales $M$, $M v$ and $M v^2$ are \emph{hard}, \emph{soft} and \emph{ultrasoft} respectively.

For simplicity, we would like to consider the situation where the mass of the scalar $m$ is
of the same order, or smaller, than the ultrasoft scale $m \siml M v^2$. In practice, this corresponds to considering Coulombic states induced by the scalar mediator, which is the regime typically studied in the existing literature (however see \eg \cite{An:2016kie,Beneke:2015lwa} for numerical studies with finite $m$).
Furthermore, should the full theory feature a scale $\Lambda$ below which perturbation theory ceases to be applicable (such as $\Lambda_\textrm{QCD}$ in strong interactions), this scale should be much smaller\footnote{In principle, it would be sufficient to demand only $\Lambda \ll M$, which would allow us to integrate out the scale $M$ perturbatively. 
The procedure of integrating out the scale $M v$ without relying on the perturbative expansion in a small coupling has been discussed in \cite{Brambilla:2000gk,Pineda:2000sz}. However, to keep the present discussion as simple as possible, we assume perturbativity at least up to scales much smaller than the bound state energy.} than $M v^2$. This ensures that the scales $M$ and $M v$ can be integrated out perturbatively.

Integrating out all degrees of freedom with energies and momenta of order $M$ and above we obtain an EFT known as the
Non-relativistic Yukawa Theory (NRY) \cite{Luke:1996hj,Luke:1997ys}. The degrees of freedom of NRY are Pauli spinor fields $\psi$ and $\chi$ describing a particle and an antiparticle respectively\footnote{To be more precise, $\psi$ annihilates a fermion, while $\chi$ creates an antifermion. This property is most easily seen in the operator approach, where the free-field Fourier decomposition of $\psi$ contains only a single particle annihilation operator $\hat{a}(\bm{p},s)$, while that of $\chi$ is proportional to the antiparticle creation operator $\hat{b}^\dagger(\bm{p},s)$.} as well as soft and ultrasoft scalar fields $\phi$. 
The Lagrangian of NRY is a double expansion in $1/M$ and $v$. While $M$ explicitly appears as a parameter in  $\mathcal{L}_{\textrm{NRY}}$, this is not the case for the velocity $v$. Therefore, to determine the velocity scaling
of the given operator it is necessary to work out power-counting rules that assign powers of velocity to the
typical operator building blocks, \ie couplings, fields and derivatives.

The fact that the energies and momenta of the $\phi$ fields can be soft or ultrasoft leads to additional complications
in the power-counting. In particular, the scaling of scalar mediators involved in potential exchanges between the heavy fermions will, in general, differ from that of the on-shell $\phi$ fields in the external states. In other words, the power counting of NRY is not homogeneous, as it has been already  discussed in \cite{Luke:1996hj,Luke:1997ys}. This is less of a problem for production and decay calculations, but turns out to be rather inconvenient when looking at the bound state properties. In section \ref{sec_pNREFT} we will show how to circumvent this problem by devising yet another EFT (pNRY) that works at energies much smaller than $M v$.

By construction, NRY is valid only at scales of order $Mv$ and below. In this energy region it must reproduce the full theory, so that both theories have identical infrared (IR) behavior. Formally, the Lagrangian of NRY contains infinitely many operators suppressed by increasing powers of $M$. This corresponds to the statement that both theories coincide in the limit $M \to \infty$. However, one can always employ the velocity scaling rules to determine which operators contribute at the given order in $v$. This is why in practice we will only need to consider a small set of relevant operators.

\subsection{Symmetries and NRY Lagrangian}
\label{sec:symm}
A crucial property of an EFT is that it must encompass the symmetries of the underlying full theory. Therefore, to construct the Lagrangian of NRY we must write down all possible operators compatible with the symmetries present in the scalar Yukawa theory. For example, each operator must be invariant under charge conjugation, parity and time reversal. Lorentz symmetry is still present in NRY, but it is not manifest.\footnote{A thorough discussion of the Poincar\'e invariance in NREFTs such as NRQCD and pNRQCD can be found in \cite{Brambilla:2001xk,Brambilla:2003nt,Vairo:2003gx,Berwein:2018fos}.} One of the implications thereof is the invariance under rotations in the 3-dimensional space. In addition to that, we will also encounter some symmetries that manifest themselves only when particles and antiparticles are treated as separate degrees of freedom and are not obvious when looking at the relativistic full theory Lagrangian.

The procedure of enumerating all operators that may appear in the given NREFT order by order in $1/M$ can be found \eg in \cite{Paz:2015uga,Gunawardana:2017zix}. This problem can be also approached using the Hilbert series framework adapted  to non-relativistic theories \cite{Kobach:2017xkw}. A more explicit way to obtain the fermion-bilinear piece of $\mathcal{L}_{\textrm{NRY}}$ is to subject the full theory Lagrangian given in \Eq\eqref{lag_mod_0} to a sequence of Foldy-Wouthuysen-Tani (FWT) transformations \cite{Foldy:1949wa,Tani:1951} or to use the equations of motion (EOM) method
as in the Heavy Quark Effective Theory (HQET) \cite{Isgur:1989vq,Isgur:1989ed,Georgi:1990um,Eichten:1989zv} (\cf \cite{Manohar:2000dt} for a pedagogical introduction to EOM). Both approaches can be iterated order by order in $1/M$ and lead to effective Lagrangians that incorporate relevant operators together with their tree-level matching coefficients. At this point it is important to stress that these techniques should not be employed mindlessly for a number of reasons. First of all, it is well-known
(\cf \eg \cite{Kinoshita:1995mt}) that FWT and EOM by construction miss all operators that are allowed by symmetries but happen to have vanishing tree-level matching coefficients.\footnote{It is clear that such operators can still become relevant at higher loop orders and hence must be included in the Lagrangian.} This should not come as a surprise, since both procedures essentially correspond to the tree-level matching. Second, the so-obtained operator basis is not guaranteed to be the most useful one and may contain redundancies. Field redefinitions can be used either to completely eliminate some of the appearing operators or to trade them for other operators. 

For example, in the case of NRQCD, NRQED or NRY one can get rid of operators with time derivative acting on the heavy fermions
by introducing suitable redefinitions of these fields. Notice that field redefinitions leave only on-shell Green functions unchanged but alter the off-shell ones. This is why the matching between the full theory and the NRY should be performed for on-shell Green functions. Nonetheless, as long as one keeps in mind the above facts, FWT and EOM can be regarded as a useful aid when working out a new NREFT containing heavy fermions. We demonstrate an explicit application of these tools to the scalar Yukawa theory up to $\mathcal{O}(1/M^2)$ in appendix \ref{app:matching}.

At $\mathcal{O}(1/M^2)$ the most general Lagrangian compatible with the symmetries of the scalar Yukawa theory can be written as 
\begin{align}
	& \mathcal{L}_{\hbox{\tiny NRY}}= 
	\nonumber \\
	& \phantom{+} \psi^\dagger \left( i \partial_0  +c_1 \,  g\phi + c_2 \frac{\bm{\nabla}^2}{2 M}  + c_{3} \frac{\phi^2}{M} +  c_{4} \frac{\phi^3}{M^2} +  c_{\hbox{\tiny D}} \frac{g\, \{\bm{\nabla} ,  \{\bm{\nabla} , \phi  \} \}}{8 M^2}  + i c_{\hbox{\tiny S}} \frac{g \sigma^i  \epsilon^{ijk} \bm{\nabla}^j \phi \bm{\nabla}^k }{4M^2}   \right) \psi 
	\nonumber \\
	&+\chi^\dagger \left( i \partial_0  + c_1' \, g \phi + c'_2 \frac{\bm{\nabla}^2}{2 M}  + c'_3 \frac{\phi^2}{M}  + c'_{4} \frac{\phi^3}{M^2}+ c'_{\hbox{\tiny D}} \frac{g\, \{\bm{\nabla} ,  \{\bm{\nabla} , \phi  \} \}}{8 M^2}  + i c'_{\hbox{\tiny S}} \frac{g \sigma^i \epsilon^{ijk} \bm{\nabla}^j \phi \bm{\nabla}^k }{4M^2}   \right) \chi \nonumber
	\\
	&+ \mathcal{L}_{\textrm{4-fermions}} 
	\nonumber
	\\
	&+\frac{d_1}{2} \partial_\mu \phi \, \partial^\mu \phi - d_2 \frac{m^2}{2} \phi^2 + \frac{ d_3}{4!} \phi^4 + \frac{d_4}{M^2}  (\partial^\mu \phi) \partial^2 (\partial_\mu \phi) + \frac{d_5}{M^2}  (\phi \partial^\mu \phi) (\phi \partial_\mu \phi)  \, , 
	\label{NREFT_lag}
\end{align}
where $\psi$ ($\chi$) is the Pauli field that annihilates (creates) a heavy fermion, while $\phi$
is the light scalar mediator. The anticommutators are defined as $\{a,b\} = a b + b a$. Furthermore, $\bm{\sigma}$ stands for Pauli matrices and we have $\partial_i = \nabla^i$. Notice that the derivatives in the bilinear fermion and antifermion sector act on all the fields (scalar and spinors) on the right. The $c_i$ and $c_i'$ are the matching coefficients of fermion and antifermion bilinears respectively, while $d_i$ belong to the scalar sector.

The $\mathcal{L}_{\textrm{4-fermions}}$ part of the Lagrangian contains 4-fermion contact interactions that describe annihilations/decays of $X \bar{X}$ pairs\footnote{$X \bar{X}$ production can be described by vacuum expectation values of 4-fermion operators containing $X \bar{X}$-Fock states between  the fermion bilinears. Such objects are therefore not included in $\mathcal{L}_{\textrm{4-fermions}}$ but will enter the corresponding production cross sections.}. These operators are necessary, since a heavy-fermion annihilation process such as  $X \bar{X} \to \phi \phi$ cannot be described via the fermion-bilinear part of the NRY Lagrangian. In this case the scalar fields must carry energies of $\mathcal{O}(M)$, yet these modes have been integrated out when constructing the NRY. This is why such processes must be described via 4-fermion interactions, where the effects of the high energy modes are incorporated in the imaginary parts of the Wilson coefficients multiplying these operators \cite{Bodwin:1994jh,Braaten:1996ix}.

The NRY Lagrangian enjoys a heavy fermion spin symmetry (HFSS) up to corrections of $\mathcal{O}(1/M^2)$, where the first spin-flipping operator shows up. It is interesting to observe that in the case of NRQCD or HQET the heavy quark spin symmetry is broken already at $\mathcal{O}(1/M)$. However, since NRY has no gauge symmetry and an operator proportional to $\bar{X} \phi \bm{\nabla} \cdot \bm{\sigma} X$ is forbidden by parity, the spin flip may occur only through an operator involving at least two spatial derivatives. The validity of the HFSS up to $\mathcal{O}(1/M^2)$ implies particularly small splittings in the spin-symmetry multiplets of $X \bar{X}$ bound states, which is an intriguing feature of the NRY phenomenology. Another symmetry of $\mathcal{L}_{\textrm{NRY}}$ that should be familiar to NREFT practitioners is the heavy fermion phase symmetry
\begin{equation}
	\psi \to e^{i \alpha} \psi, \quad  \chi \to e^{i \beta} \chi, \quad \alpha,\beta \in \mathbb{R},
\end{equation}
which implies separate conservation of the number of particles and antiparticles.

\subsection{Power counting}
\label{sec:power}
To derive the power-counting rules of the theory we can make use of the standard arguments\footnote{Strictly speaking, these
	argument are rigorous only in the context of non-relativistic quantum mechanics and must be revised for a theory that features a non-perturbative regime.} used in NRQCD \cite{Bodwin:1994jh}. To this end it is useful to employ a quantum mechanical perspective before the second quantization, where we can interpret $\psi$ as a wave function interacting with an external potential $\phi$. The wave function normalization condition 
\begin{equation}
\int d^3 x \, \psi^\dagger(\bm{x},t) \psi(\bm{x},t) = 1,
\end{equation}
together with our previous estimate of the typical bound state radius $r \sim 1/{Mv}$ readily suggests that $\int d^3 x \sim 1/(Mv)^3$ and therefore $\psi \sim (M v)^{3/2}$. A spatial derivative acting on $\psi$ probes its typical 3-momentum, so that $\bm{\nabla }\psi \sim Mv \, \psi$. The equation satisfied by $\psi$ at the lowest order in the $1/M$ expansion reads 
\begin{equation}
 \left( i \partial_0  + \frac{\bm{\nabla}^2}{2 M}  - g\phi  \right) \psi = 0,
 \label{sch_nry}
\end{equation}
where we have anticipated the tree-level results for the matching coefficients $c_1$ and $c_2$ (cf.~\Eq\eqref{tree_level_bilinear}). Here $g \phi$ plays the role of the leading-order contribution to the interacting part of the quantum mechanical Hamiltonian. Using the virial theorem for bound states we can estimate that $\partial_0 \psi \sim Mv^2 \psi$ and $g \phi \sim M v^2$. The same argument applies also to the scaling of the $\chi$ field. In a similar manner, we may also pass to a picture in which the wave function $\phi$ satisfies the following Klein-Gordon-Schr\"odinger equation (again at  lowest order in $1/M$)
\begin{equation}
	(\partial_0^2 - \bm{\nabla}^2) \phi + m^2 \phi - \frac{\lambda}{3!} \phi^3 - g \bar{X} X = 0,
\end{equation}
where the last term scales as $g M^3 v^3$. If we assume that the typical momentum of $\phi$ scales as $M v$, then the virial theorem implies that $g Mv \sim \phi$. Hence, $g^2  \sim v$ for a Coulombic state and $\phi \sim M v^{3/2}$. Notice also that
$\lambda \phi^3 \sim M^3 v^{7/2}$ may seem much less suppressed than $(g \phi)^3 \sim M^3 v^6$. However, since $\lambda$ does not appear in the fermion-bilinear part of the NRY Lagrangian, a diagram involving $\lambda$ must also contain at least one insertion of $g \phi$ that couples directly to the fermion current. This accounts for an extra suppression of processes involving the scalar self-coupling with $\lambda$.

Notice also that if the energy and momentum of $\phi$ scale as $M v^2$, we find $ v^4 \phi \sim g M v^3$ and consequently $g^2 \sim v^3$, upon using $g \phi \sim M v^2$. In this case we would actually need less operators to describe the same observable at the given order in $v$ as compared to the previous counting. Yet, to be on the safe side, in the following we will adopt the more conservative counting with $\bm{\nabla} \phi \sim M v \phi $. We summarize the scaling rules in figure~\ref{fig:NRYT_scales}.

\section{Applications of NRY to dark matter}
\label{sec_NREFT}
In this section we adapt the general discussion of section \ref{sec:NREFTs_general} to the DM phenomenology, and derive the matching coefficients of the low-energy version of the model Lagrangian \Eq\eqref{lag_mod_0}, namely the parameters of the NRY (\ref{NREFT_lag}). The effective Lagrangian comprises unknown coefficients that have to be fixed by the matching procedure. In practice, one  computes on-shell Green's functions in the full theory in \Eq\eqref{lag_mod_0} and in the effective theory and demands their equality at a matching scale $\mu_{\hbox{\scriptsize match}}$ with $m , M v^2, Mv \ll \mu_{\hbox{\scriptsize match}} \ll M$. The relative size of the smaller scales is irrelevant here. Through the matching coefficients, which could be obtained at arbitrary loop order, the low-energy theory is also organized as an expansion in the couplings $g$ and $\lambda$. As it is common in DM models, we assume the scalar self-coupling to satisfy  $\lambda \sim g^2$,
which facilitates the organization of the perturbation series. Furthermore, we define $\alpha \equiv g^2/4 \pi$ to organize the power counting of the low-energy theories. In this work $\lambda$ will barely play any role.

A non-relativistic regime for dark particles is relevant both for annihilations during the thermal freeze-out, as well as in the present-day galactic halos. In the latter case, typical DM velocities are of order $10^{-4}$-$10^{-3}$ in units of $c$, \cf \eg \cite{Kuhlen:2009vh,Kuhlen:2013tra}. In the former case, DM particles are kept in chemical equilibrium through  interactions  with  the  thermal  bath  until $T \ll M$ and  gradually  freeze  out at temperatures $T \sim M/25$.\footnote{The chemical freeze-out temperature $T$ can be estimated by equating the expansion and annihilation rates $H \sim n_{{\rm{eq}}} \langle \sigma_{\hbox{\scriptsize ann}} v_{\hbox{\scriptsize rel} } \rangle$, namely $T^2/M_{\hbox{\tiny Pl}} \sim \left( \frac{MT}{2\pi}\right)^{3/2} e^{-M/T} \alpha^2/M^2$,  where $H$ is the Hubble rate of the Universe, $\alpha$ is some fine structure constant and $M_{\hbox{\tiny Pl}} \simeq 1.2 \times 10^{19}$ GeV. After the chemical equilibrium is lost, kinetic equilibrium is usually kept for longer times and it provides a thermal distribution for the DM momenta (and velocities).}  Annihilations continue even during later stages where the DM particles are still in kinetic equilibrium. In this situation most of the energy of a DM particle is sourced by its mass and, for non-relativistic species, the typical momentum is $|\bm{p}|=\sqrt{T M}=M\sqrt{T/M}$. One usually identifies an average velocity $v \approx  \sqrt{T/M}$, which is smaller than unity in the regime of interest. For above-threshold particle-antiparticle pairs feeling a Coulomb-like potential, the regime $v \sim \alpha$ signals the potential energy $E_{\textrm{pot}} \sim \alpha /r \sim M \alpha v$ being the same as the kinetic energy $M v^2$ \cite{Kim:2016kxt}. For $X \bar{X}$ in a bound state, the velocity estimate of the relative motion is fixed at $v \sim \alpha$. In a perturbative regime, this again gives a velocity smaller than unity. Since the temperature of the plasma is $T \ll M$, the temperature scale is treated on the same footing with other smaller scales, and does not affect neither the matching and nor the form of the NRY (\cf discussions in \cite{Escobedo:2008sy,Brambilla:2008cx} for NRQED and NRQCD at finite temperature and \cite{Biondini:2013xua} for an explicit derivation of an NREFT for Majorana fermions in a thermal bath). 

In summary, at energies much smaller than $M$, the degrees of freedom are non-relativistic Dirac fermions and antifermions, including bound states and near-threshold states, and scalars with energies and momenta much smaller than $M$. 
The NRY presented in \Eq\eqref{NREFT_lag} is then a suitable field theory that describes non-relativistic DM particles and their dynamics. 
The first two lines of \Eq\eqref{NREFT_lag} encode interactions between the non-relativistic fermion (and antifermion) and the light scalar mediator. An important difference with respect to NRQED/NRQCD is the lack of gauge symmetry. Hence, the effective Lagrangian \Eq\eqref{NREFT_lag} contains no covariant derivatives and the form of effective operators containing scalars and fermions is not constrained accordingly. We discuss the matching of the bilinear sector in section \ref{sec_NREFT_matching_bil}. 
The last line of \Eq\eqref{NREFT_lag} comprises 4-fermion operators, which account for DM pair annihilations in the low-energy theory. The corresponding cross section $X \bar{X} \to \phi \phi$ is a key ingredient for the determination of the relic density governed by the freeze-out as well as present day annihilations in the Milky-Way. In this work, we do not consider pair annihilations induced by interactions in $\mathcal{L}_{\textrm{portal}}$. We address the 4-fermion operator in section \ref{sec_NREFT_application_four_f}.

\subsection{NRY Matching}
\label{sec_NREFT_matching}
We now discuss the derivation of the matching coefficients of the low-energy theory given in \Eq\eqref{NREFT_lag}. As already anticipated, this procedure amounts to enforcing the equality of on-shell scattering amplitudes in the full theory (\ref{lag_mod_0}) with on-shell scattering amplitudes constructed with the general expressions of the NRY in terms of $\psi$, $\chi$ and $\phi$ and the unknown matching coefficients. The matching scale provides a UV cut-off for the low-energy theory, above which NRY is not reliable. Clearly, the fundamental theory and the NREFT have a different UV behavior, whereas the infrared (IR) properties are the very same. Only high-energy modes of order $M$ (integrated out in NRY)  contribute to the matching coefficients of \Eq\eqref{NREFT_lag}. In other words, when computing scattering amplitudes there can be residual IR contributions that are not included in the matching coefficients, because they appear on both sides of the matching condition. 

Most of the calculations done in the course of this work (\eg determination of the matching coefficients, 
derivation of the Feynman rules, manipulations of the EFT Lagrangians etc.) were carried out not only by pen and paper but also using software tools for automatic calculations. For the latter we employed \textsc{Mathematica} packages \textsc{FeynArts} \cite{Hahn:2000kx}, \textsc{FeynRules} \cite{Alloul:2013bka} and \textsc{FeynCalc} \cite{Mertig:1990an,Shtabovenko:2016sxi,Shtabovenko:2020gxv}. The automation of non-relativistic calculations was significantly
simplified by making use of \textsc{FeynCalc} 9.3 and the \textsc{FeynOnium} \cite{Brambilla:2020fla} extension that allow for algebraic manipulations of Cartesian tensors and Pauli matrices. Furthermore, an experimental interface to \textsc{QGRAF} \cite{Nogueira:1991ex} diagram generator was added to the development version of the \textsc{FeynHelpers} \cite{Shtabovenko:2016whf} extension. This allowed us to generate Feynman diagrams for non-relativistic EFTs in a straightforward fashion. All new functions that were developed while working on this project should be made publicly available and properly documented in the upcoming versions of \textsc{FeynCalc}, \textsc{FeynOnium} and \textsc{FeynHelpers}.

\subsubsection{Fermion bilinear and scalar sector}
\label{sec_NREFT_matching_bil}
\begin{figure}[t!]
    \centering
    \includegraphics[scale=0.42]{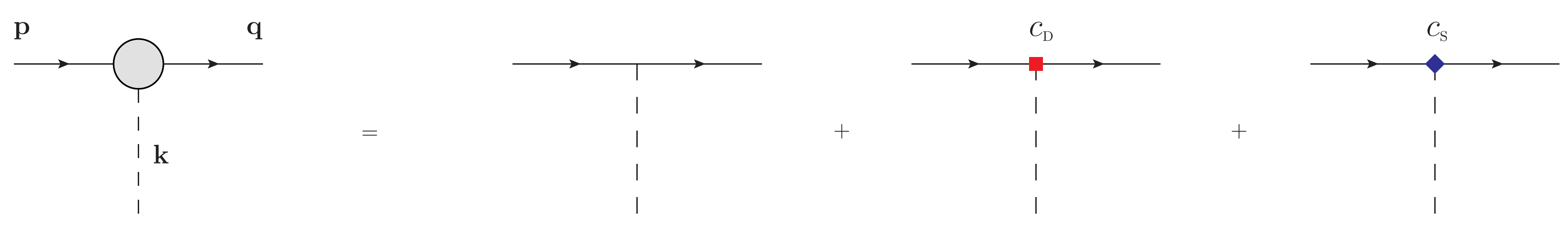}
    \caption{Matching of the effective vertices with an incoming and an outgoing fermion and one scalar. The blob on the diagram on the left of the equality indicates the possibility of including quantum corrections in the fundamental theory \Eq\eqref{lag_mod_0}. Here the matching is done at tree-level. The diagrams on the right hand side correspond to the effective vertices of the NRY with the corresponding coefficients fixed by the matching.}
    \label{fig:Fer_sca_match}
\end{figure}

Let us discuss the matching coefficients of the bilinear fermion and antifermion sectors, first two lines in \Eq\eqref{NREFT_lag}. This amounts to comparing scattering amplitudes with one incoming and one outgoing fermion and scalar mediators (one, two or three of the latter field). The diagrammatic representation of the matching for a fermion interacting with one single scalar field is given in figure \ref{fig:Fer_sca_match}. In this work, we consider the matching of the NREFT Lagrangian at tree-level as far as the fermion (antifermion) bilinear is concerned. For the trilinear coupling this means that it is sufficient to work at order $g$. However, we remark that this procedure is general and applicable to the matching at any loop order. 

We collect some details in the appendix \ref{sec:matching}, whereas here we list the results for the matching coefficients that read 
\begin{eqnarray}
    && c_1=-c_1'=-1 \, , \quad c_2=c_2'=1\, ,  \quad  c_{\hbox{\tiny D}}=-c'_{\hbox{\tiny D}}=-1 \,, 
    \nonumber 
    \\
    &&c_{\hbox{\tiny S}}=-c'_{\hbox{\tiny S}}=-1 \, , \quad  c_3=c_4=c_3'=c_4'=0 \,.
    \label{tree_level_bilinear}
\end{eqnarray}
For $c_3, c_4, c_3'$ and $c_4'$, we have considered the matching of diagrams with two and three external scalars respectively. Consistently with the findings from the FWT and EOM methods (\cf appendix \ref{sec:matching}), these matching coefficients are found to vanish at tree-level.

The matching coefficients $c_1,c_{\hbox{\tiny D}},c_{\hbox{\tiny S}}$ may receive $\mathcal{O}(g^2)$, $\mathcal{O}(\lambda)$ corrections (not addressed in this work\footnote{According to the power counting given in section \ref{sec_pNREFT}, the matching coefficients in the bilinear sector beyond tree-level are needed to compute the bound-state spectrum at order $M \alpha^5$, as well as corrections to bound-state formation rates, which is beyond the scope of this work.}), whereas $c_3,c_4$ may start getting non-trivial contributions at one-loop level. The coefficients of the kinetic terms $c_2$ and $c_2'$ are fixed to unity to all orders in perturbation theory owing to the  reparametrization invariance.

There is an important aspect we want to highlight. As one may read off from \Eq\eqref{tree_level_bilinear}, there is a relative sign difference between the particle and antiparticle interactions with the scalar field. At order  $\mathcal{O}(1/M^0)$ this is in contrast to the situation in NRQED and NRQCD, where the signs are the same. This very difference will be the reason behind the appearance of monopole and quadrupole interactions in the lower energy EFT that we will derive in section \ref{sec_pNREFT}, instead of typical dipole interactions of pNRQED and pNRQCD.

As for the scalar sector described in third line of \Eq\eqref{NREFT_lag}, we equally perform the matching at tree-level only. Our guidance here is again the power counting of the pNRY that will be given in section \ref{sec_pNREFT}. Postponing the one-loop matching of the NRY to a future work on the subject, one can simply obtain the matching coefficients at tree-level to be 
\begin{equation}
    d_1 =1 \, \quad d_2 =1 \, \, \quad d_3 =-\lambda \, , \quad d_4=d_5 =0 \, . 
\end{equation}

\subsubsection{Four-fermion operators and annihilation cross section}
\label{sec_NREFT_application_four_f}
\begin{figure}[t!]
\centering
\includegraphics[scale=0.52]{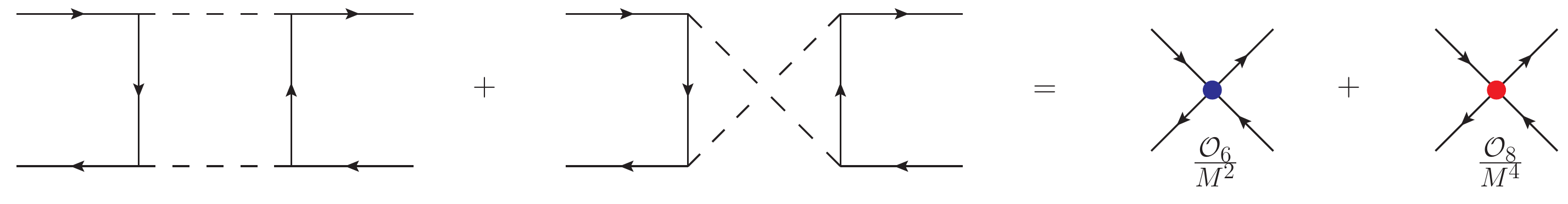}
\caption{\label{fig:DM_0_ann_sca} Diagrammatic matching between the relativistic theory (diagrams on the l.h.s) and the corresponding four-particle local interactions in the NREFT (diagrams on the r.h.s). The latter correspond to the dimension-6 and dimension-8 operators of the four-fermion sector (\ref{NREFT_lag}) respectively.}
\end{figure}
As anticipated, the NRY can readily describe heavy pair annihilations in terms of local 4-fermion operators in \Eq\eqref{NREFT_lag}. The inclusive annihilation rate can be recast in terms of an amplitude that conserves the number of the heavy particles by means of the optical theorem: the imaginary part of the loop amplitude with four external heavy fermion legs is related to the cross section of the process $X \bar{X} \to \phi \phi$ \cite{Bodwin:1994jh,Braaten:1996ix}, \cf figure \ref{fig:DM_0_ann_sca}. For the model at hand, it is known that the annihilation cross section is velocity suppressed \cite{Pospelov:2007mp}. This will be reflected in a vanishing contribution from the velocity- (or derivative-) independent operators, that are of dimension-6. They read \cite{Bodwin:1994jh}  
\begin{eqnarray}
( \mathcal{L}_{\textrm{4-fermions}} )_{d=6} = \frac{f(^1S_0)}{M^2} \psi^\dagger \chi \, \chi^\dagger \psi + \frac{f(^3S_1)}{M^2} \psi^\dagger \, \bm{\sigma} \, \chi \cdot \chi^\dagger \, \bm{\sigma} \, \psi
\label{dimension_6_lag}
\end{eqnarray}
The spectroscopy notation is borrowed from NRQED/NRQCD, so that one can classify  the annihilations in terms of the total spin $S$ of the pair, the relative angular momentum $L$ and the total angular momentum $J$, by writing  $^{2S+1}L_J$.
 Then, we consider dimension-8 operators, which comprise higher powers in $1/M$. These are compensated by derivatives acting on the fermion (antifermion) fields, that induce  velocity suppressed contributions due to $\bm{\nabla} \psi \sim Mv$ (\cf section \ref{sec:NREFTs_general}). As we shall see, they provide the leading contribution to the annihilation process $X \bar{X} \to \phi \phi$. Of course, higher dimensional operators (further suppressed in the velocity expansion) are allowed as well but will not be considered in this work. The explicit structure of  the dimension-8 operators in \Eq\eqref{NREFT_lag} reads \cite{Bodwin:1994jh}
\begin{eqnarray}
( \mathcal{L}_{\textrm{4-fermions}} )_{d=8} &=& \frac{f(^1P_1)}{M^4} \mathcal{O}(^1P_1) + \frac{f(^3 P_0)}{M^4} \mathcal{O}(^3 P_0) + \frac{f(^3 P_1)}{M^4} \mathcal{O}(^3 P_1)  \nonumber 
\\
 &+& \frac{f(^3 P_2)}{M^4} \mathcal{O}(^3 P_2) + \frac{g(^1 S_0)}{M^4} \mathcal{P}(^1 S_0) + \frac{g(^3 S_1)}{M^4} \mathcal{P}(^3 S_1)
 \nonumber 
 \\
  &+& \frac{g(^3 S_1,^3 D_1 )}{M^4} \mathcal{P}(^3 S_1, ^3 D_1) + \cdots 
\label{dimension_8_lag}
\end{eqnarray}
where the operators explicitly included are  
\begin{eqnarray}
\mathcal{O}(^1P_1) &=& \psi^\dagger \left( - \frac{i}{2} \overset{\leftrightarrow}{\bm{\partial}} \right) \chi \cdot \chi^\dagger \left( -\frac{i}{2} \overset{\leftrightarrow}{\bm{\partial}} \right) \psi \, , 
\\
 \mathcal{O}(^3P_0) &=& \frac{1}{3} \psi^\dagger \left( -\frac{i}{2} \overset{\leftrightarrow}{\bm{\partial}} \cdot \bm{\sigma} \right) \chi\,  \chi^\dagger \left( -\frac{i}{2} \overset{\leftrightarrow}{\bm{\partial}} \cdot \bm{\sigma} \right) \psi\, , 
 \\
 \mathcal{O}(^3P_1) &=& \frac{1}{3} \psi^\dagger \left( -\frac{i}{2} \overset{\leftrightarrow}{\bm{\partial}} \times \bm{\sigma} \right) \chi  \cdot  \chi^\dagger \left( -\frac{i}{2} \overset{\leftrightarrow}{\bm{\partial}} \times \bm{\sigma} \right) \psi\, , 
 \\
 \mathcal{O}(^3P_2) &=&  \psi^\dagger \left( -\frac{i}{2} \overset{\leftrightarrow}{\partial}  \phantom{s}^{(i} \sigma^{j)} \right) \chi\,  \chi^\dagger \left( -\frac{i}{2} \overset{\leftrightarrow}{\partial}  \phantom{s}^{(i} \sigma^{j)} \right) \psi \, ,
 \\
 \mathcal{P}(^1 S_0) &=& \frac{1}{2} \left[ \psi^\dagger \chi \, \chi^\dagger \left( \frac{i}{2} \overset{\leftrightarrow}{\bm{\partial}} \right)^2 \psi + \textrm{h.c.} \right] \, ,
  \\
 \mathcal{P}(^3 S_1) &=& \frac{1}{2} \left[ \psi^\dagger \bm{\sigma} \chi \, \cdot \chi^\dagger \bm{\sigma} \left( \frac{i}{2} \overset{\leftrightarrow}{\bm{\partial}} \right)^2 \psi + \textrm{h.c.} \right] \, ,
 \\
 \mathcal{P}(^3 S_1, ^3 D_1) &=& \frac{1}{2} \left[ \psi^\dagger \sigma^i \chi \, \cdot \chi^\dagger \sigma^j  \left( \frac{i}{2} \right)^2 \overset{\leftrightarrow}{\partial}  \phantom{x}^{(i} \overset{\leftrightarrow}{\partial}\,^{j)}  \psi + \textrm{h.c.} \right] \, , 
\end{eqnarray}
where $\overset{\leftrightarrow}{\bm{\partial}}$  is the difference between the derivative acting on the spinor to the right \emph{and} on the spinor to the left, namely $\chi^\dagger \overset{\leftrightarrow}{\bm{\partial}} \psi \equiv \chi^\dagger (\bm{\partial} \psi)- (\bm{\partial} \chi)^\dagger \psi$. The notation $T^{(ij)}$ for a rank 2 tensor stands for its traceless symmetric components, $T^{(ij)}=(T^{ij}+T^{ji})/2-T^{kk}\delta^{ij}/3$. As pointed out in \cite{Bodwin:1994jh}, we may also have operators with the derivative acting on the product of the spinor fields $\psi^\dagger$ and $\chi$ or $\chi^\dagger$ and $\psi$. The matrix elements of such operators are proportional to the total momentum of the pair $X \bar{X}$, that is zero in the rest frame of the particle-antiparticle pair.

The detailed derivation of the matching coefficients can be found in appendix \ref{sec:matching}, whereas here we merely list the results 
\begin{eqnarray}
   && {\rm{Im}}[f(^1S_1)] =  {\rm{Im}}[f(^3S_1)] = 0 \, ,
   \\
   && {\rm{Im}}[f(^1P_1)] =  {\rm{Im}}[f(^3P_1)] = 0  \, ,
   \\
   &&   {\rm{Im}}[f(^3P_0)] = \frac{25}{6} \pi \alpha^2  \, , \quad  {\rm{Im}}[f(^3P_2)] = \frac{1}{15} \pi \alpha^2  \, ,
   \\
   && {\rm{Im}}[g(^1S_0)] =  {\rm{Im}}[g(^3S_1)]  = {\rm{Im}}[g(^3S_1, ^3D_1)] = 0 \, .
    \label{match_sca_dim_8}
\end{eqnarray}
Both matching coefficients of the dimension-6 operators in \Eq\eqref{dimension_6_lag} are zero at the order we are working. Accordingly, they do not contribute to the pair-annihialtions. We observe that the annihilating fermions are always in the spin triplet configuration with the orbital angular momentum $L=1$
and definite total angular momentum values $J=0,2$. The allowed combinations of $L,S$, and then $J$, are constrained by the symmetry of the fundamental Lagrangian \Eq\eqref{lag_mod_0}, that are also inherited by the NRY \Eq\eqref{NREFT_lag}. Since the scalar does not carry any spin, the conservation of parity and the total angular momentum forbids $S=0$, and impose $\Delta L=0$ or even.

We conclude this section by reproducing the non-relativistic annihilation cross section for the process $X \bar{X} \to \phi \phi$. In order to compare to the results in the literature, we average over spin polarizations of the incoming fermion and antifermion. The cross section then reads
\begin{eqnarray}
    \sigma_{ \textrm{ann}} =  \frac{2{\rm{Im}}[\mathcal{M}_{\hbox{\tiny NR}}(\psi\chi \to \psi\chi)]}{4 \times 2 v } \, ,
    \end{eqnarray}
    where we used the non-relativistic flux factor with the relative velocity in the center of mass frame, so that $v_{\textrm{rel}}=|\bm{v}_\psi-\bm{v}_\chi|=2 v$. The imaginary part of the non-relativistic amplitude $\mathcal{M}_{\hbox{\tiny NR}}$ can be readily computed using the Lagrangian from \Eq\eqref{dimension_8_lag} and the matching coefficients \Eq\eqref{match_sca_dim_8}, and upon setting $\bm{v}=\bm{v}'$ in \Eq\eqref{NREFT_t_and_u} (since we consider the scattering amplitude of $\psi\chi \to \psi\chi$) we obtain
    \begin{eqnarray}
  \sigma_{\textrm{ann}} v_{\textrm{rel}} = \frac{  {\frac{2}{3}{\rm{Im}}[f(^3P_0)]} + {\frac{10}{3}{\rm{Im}}[f(^3P_2)]}  }{ 8 M^2}v_{\textrm{rel}}^2 = \frac{3 \pi \alpha^2}{8M^2} v_{\textrm{rel}}^2  \, .
    \label{NR_hard_cross_section_S}
\end{eqnarray}
The result agrees with the literature, \cf \eg \cite{Wise:2014jva,An:2016kie}.

As was pointed out in \cite{Petraki:2015hla}, and can be inferred from the benchmark point used in \cite{An:2016kie},
large values of $\alpha$ in this model can be of particular phenomenological interest. It is in the reach of the NRY, and subject of another work \cite{Vlady_and_Simo_2}, to derive such higher order corrections in the matching coefficients of the bilinear and 4-fermion sector, and to inspect their impact on, \eg the annihilation cross section.

\section{Potential non-relativistic Yukawa theory}
\label{sec_pNREFT}
In the previous section, we integrated out the hard degrees of freedom with energies of order $M$ as well as fermion/antifermion fluctuations of the same order. Here we want to integrate out soft degrees of freedom with energies of the order of the relative momentum of the pair $\bm{p} \sim M v$.\footnote{A distinction between potential photons, \ie photons with $k^0 \sim M \alpha^2$ and $k \sim M \alpha$ from
soft photons, \ie photons with $k^0 \sim k \sim M \alpha$, can also be considered \cite{Beneke:1997zp,Griesshammer:1998tm} and this could apply to the scalar mediator as well. This distinction is not that relevant in the formulation we are following
since, as done for the QED case, both potential and soft photons are integrated out at the same time when matching NRQED to
pNRQED.} The corresponding effective theory takes the form of a pNREFT and we can rely on the techniques employed in the derivation of pNRQED as long as we assume that the mass of the scalar satisfies $m \ll M v$. This condition implements a Coulomb-like regime and implies the scaling $v \sim \alpha$ for the velocity. Moreover, we are allowed to treat the scalar mediator as effectively massless in the matching between the NRY and the pNRY, upon relying on the scale hierarchy $M \alpha \gg M\alpha^2, m$, irrespective of the relative size of the smaller scales. We comment on the case $m \sim M \alpha$ later in this section. 

Let us come to the construction of the pNRY Lagrangian. First of all, as the two-point functions are not sensitive to the relative momentum of the pair, the fermion bilinears of the NRY from \Eq\eqref{NREFT_lag} and the pNRY will look the same. That said, one has to keep in mind that only scalar fields with ultrasoft momenta are kept in the latter EFT. Conversely, diagrams with four-fermion external legs are sensitive to the relative momentum and non-trivial contributions will be generated: they are the \emph{potential} terms in the pNREFT Lagrangian \cite{Pineda:1998kj,Pineda:1998kn}. The important point to be stressed here is that the appearance of the potential terms can be seen as the effect of integrating out soft scalars, and hence the potential can be extracted by matching the NRY to the pNRY.

In order to elucidate on the distinction between soft and ultrasoft scalars, and to introduce the degrees of freedom of pNRY, we project the NRY onto the particle-antiparticle sector as follows
\begin{equation}
\int d^3 \bm{x}_1 d^3 \bm{x}_2 \varphi_{ij}(t,\bm{x}_1, \bm{x}_2) \psi^\dagger_i (t,\bm{x}_1) \chi_j (t,\bm{x}_2) | \phi_{\hbox{\tiny US}}\rangle \, ,
\label{proj_fock}
\end{equation}  
where $i,j$ are spin indices, while the state $| \phi_{\hbox{\tiny US}}\rangle$ contains no heavy particles/antiparticles and an arbitrary number of scalars with energies much smaller than $M\alpha$. Here, $\varphi_{ij}(t,\bm{x}_1, \bm{x}_2)$ is a wave function
representing the $X\bar{X}$ system. After the projection it will be eventually promoted to a bi-local field. As a next step, one recognizes the relative distance of the pair $\bm{r}=\bm{x}_1-\bm{x}_2$ to have typical size of the inverse of the soft scale $M \alpha$, or the inverse Bohr radius of a Coulombic state, \cf \Eq\eqref{Coulomb_en_levels}. This can be considered a small scale as compared to the typical wave-length of the ultrasoft scalars, which is of the order of the inverse of $M \alpha^2$. According to the projection (\ref{proj_fock}), scalar fields now appear at points $\bm{x}_1$ and $\bm{x}_2$, and we can ensure that they are ultrasoft by expanding them about the center of mass coordinate $\bm{R}=(\bm{x}_1+\bm{x}_2)/2$ upon using $\bm{r} \ll \bm{R}$.

One has to evaluate the leading order interaction between particle-antiparticle pairs and the ultrasoft scalar field, namely the combination $g (\phi(\bm{x}_1)+\phi(\bm{x}_2)) $, around the relative coordinate up to  $\mathcal{O}(r^2)$ as follows 
 \begin{eqnarray}
     \phi(\bm{x}_1) &=& \phi\left( \bm{R}+\frac{\bm{r}}{2} \right) \simeq \phi(\bm{R}) + \frac{\bm{r}}{2} \cdot \bm{\nabla} \phi (\bm{R}) + \frac{1}{8} r^i r^j \nabla_R^i \nabla_R^j \, \phi (\bm{R})\, , \\
    \phi(\bm{x}_2) &=&  \phi\left(  \bm{R}-\frac{\bm{r}}{2} \right) \simeq \phi(\bm{R}) - \frac{\bm{r}}{2} \cdot \bm{\nabla} \phi (\bm{R})  + \frac{1}{8} r^i r^j \nabla_R^i \nabla_R^j \, \phi (\bm{R}).
 \end{eqnarray} 
 As a consequence, the dipole terms, namely the ones linear in $r$, cancel exactly. This is a peculiar feature of the Yukawa-type theory \Eq\eqref{lag_mod_0} and its low-energy version \Eq\eqref{pNREFT_sca_0} which distinguishes them from pNRQED (and pNRQCD) where dipole transitions naturally arise.
In summary, we write the Lagrangian in terms of the wave function field $\varphi(\bm{r},\bm{R},t)$ and the ultrasoft scalars $\phi(\bm{R},t)$ as follows
\begin{eqnarray}
     L_{\hbox{\tiny pNRY}}  &=& \int d^3 \bm{r} \, d^3 \bm{R}  \, \varphi^\dagger(\bm{r},\bm{R},t) \left\lbrace  i \partial_0 +\frac{\bm{\nabla}^2_{\bm{r}}}{M} +\frac{\bm{\nabla}^2_{\bm{R}}}{4M} + \frac{\bm{\nabla}^4_{\bm{r}}}{4 M^3} - V(\bm{p},\bm{r},\bm{\sigma}_1,\bm{\sigma}_2)\right. 
     \nonumber \\
     &&\hspace{3.1 cm} \left. - 2 g \phi(\bm{R},t) -g\frac{ r^i r^j }{4}  \left[ \nabla_R^i \nabla_R^j \, \phi (\bm{R},t)   \right] -   g \phi(\bm{R},t) \frac{\bm{\nabla}^2_{\bm{r}}}{M^2}    \right\rbrace \varphi(\bm{r},\bm{R},t)
     \nonumber
     \\
     &+& \int d^3 \bm{R} \left[ \frac{d_1}{2} \partial^\mu \phi \, \partial_\mu \phi - d_2 \frac{m^2}{2} \phi^2 + \frac{ d_3}{4!} \phi^4 + \frac{d_4}{M^2}  (\partial^\mu \phi) \partial^2 (\partial_\mu \phi) + \frac{d_5}{M^2}  (\phi \partial^\mu \phi) (\phi \partial_\mu \phi)  \right]\, ,
     \nonumber
     \\
     \label{pNREFT_sca_0}
 \end{eqnarray}
 where the square brackets in the second line of \eqref{pNREFT_sca_0}
 indicate that the spatial derivatives act on the scalar field only, which has to be understood as multipole expanded in the last line of \Eq\eqref{pNREFT_sca_0} as well. To avoid cluttering the notation we suppress the spin indices of the bilocal fields that are contracted with each other.
 
 At variance with NRY \eqref{NREFT_lag}, each term in the pNRY has a well-defined scaling: $\partial_0 \sim M \alpha^2$, the inverse relative distance and the corresponding derivative   ${\bm{r}}^{-1}, \bm{\nabla}_{\bm{r}} \sim M \alpha$, whereas the scalar field and the center-of-mass derivative $g\phi, \bm{\nabla}_{\bm{R}} \sim M \alpha^2$.
 The potential is understood as a matching coefficient and is organized as an expansion in $\alpha(M)$ and $\lambda(M)$, as well as $1/M$ (inherited from NRY), the  coupling $\alpha(1/r)$ and the relative distance $r$ (as proper of pNRY). In the following, we parametrize it as $V = V^{(0)} + \delta V$. In the case of $m \ll M \alpha$ the leading order potential reads $V^{(0)}=-\alpha(1/r)/r$, which is a Coulomb potential. It is important to specify  the scale at which $\alpha$ is evaluated, since it helps to keep track of the matching between the full theory from \Eq\eqref{lag_mod_0} and the NRY from \Eq\eqref{NREFT_lag}, and between the latter with pNRY given in \Eq\eqref{pNREFT_sca_0}. The corrections to the potential $\delta V$ are needed to compute observables, such as the energy spectrum, at next-to-leading order. Here, we aim at extracting the potential at the first non-trivial order $M \alpha^4$, as an application of pNRY and the corresponding power counting. We discuss the potential matching in the following section \ref{sec_pNREFT_matching}. As for the kinetic terms, we included contributions up to order $M \alpha^4$ as well.
 
 Next, in the second line of \Eq\eqref{pNREFT_sca_0}, we see the appearance of a monopole and a quadrupole interactions as well as interactions involving the derivative in the relative distance. Such structures are found by performing the so-called \emph{multipole expansion} of the ultrasoft fields, here the scalar mediator, and terms up to order $M \alpha^4$ are retained. 
 The absence of dipole transitions in this model was already pointed out in \cite{Wise:2014jva,An:2016kie} when dealing with the calculation of the bound-state formation. In our approach, the absence of such terms is already manifest at the level of the Lagrangian, where the degrees of freedom and their interactions are spelled out at the energy scale of interest for bound-state calculations. These are the wave function field of the particle-antiparticle pair and ultrasoft scalars, that interact via monopole and quadrupole interactions. The term with the derivative $\nabla_{\bm{r}}$ in the second line of \Eq\eqref{pNREFT_sca_0} arises from the $1/M^2$ spin-independent operator in \Eq\eqref{NREFT_lag}, and is of the same order as the quadrupole term, namely $M \alpha^4$. Indeed, 
the contributions of spin-dependent operators are subleading in the power counting. The presence of the ultrasoft scalar-derivative coupling conforms with the findings of \cite{Wise:2014jva}.\footnote{The authors of \cite{Wise:2014jva} do not use EFT methods and hence provide no power-counting rules. The coupling $g \phi (\bm{R},t)\nabla^2_{\bm{R}} /M^2 $ has been considered in the non-relativistic Hamiltonian, which is however $\alpha^2$ suppressed with respect to $g \phi (\bm{R},t)\nabla^2_{\bm{r}} /M^2 $.} 
 
\subsection{Potential matching}
\label{sec_pNREFT_matching}
\begin{figure}[t!]
     \centering
     \includegraphics[scale=0.65]{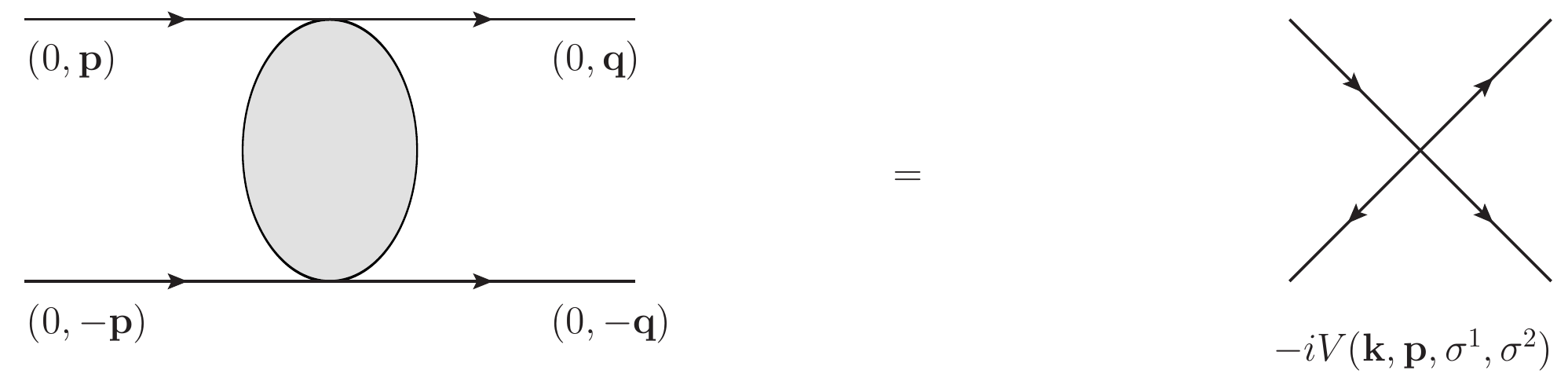}
     \caption{Green's function in NREFT and pNREFT, where the momentum assignment is explicitly shown. The momentum transfer is $\bm{k}=\bm{p}-\bm{q}$, where $\bm{k}$ is the momentum carried by the scalar mediator. The potential depends on $\bm{p}$ and $\bm{k}$ and the spin of the particle and antiparticle. }
     \label{fig:potential_matching}
 \end{figure}

In this section we address the matching between NREFT and pNREFT. This procedure brings us to the systematic derivation of the particle-antiparticle potential, which can be understood as a matching coefficient of the pNRY. This procedure is rather general, and it has been generalized at finite temperature in \cite{Escobedo:2008sy,Escobedo:2010tu} for pNRQED, as well as for pNRQCD \cite{Brambilla:2008cx}: any scale larger than $M \alpha^2$ may contribute to the potential. For the moment, let us assume that the scalar mass is much smaller than the soft scale $M \alpha$ and can, therefore, be treated on the same footing with the ultrasoft scale $M \alpha^2$ in the matching. At variance with on-shell Green's functions exploited in the matching between the relativistic theory \Eq\eqref{lag_mod_0} and NRY \Eq\eqref{NREFT_lag}, here we enforce the equality between off-shell four-fermion Green's functions. The reason is that this is the typical situation of particles in a bound state or near-threshold. We give a diagrammatic representation of the potential matching in figure \ref{fig:potential_matching}. Then, external momenta of the fermions are soft $\sim M \alpha$, whereas the energies are at the ultrasoft scale $M \alpha^2$, so that one can expand in such a scale. For the sake of the matching performed here, the ultrasoft scale  can be put to zero. This amounts to simplifying the scalar propagator to $-i/ \bm{k}^2$ and taking the fermion propagators in loop diagrams of the NREFT side of the matching as static \cite{Manohar:1997qy,Pineda:1998kj}. Loop diagrams on the pNREFT side vanish in dimensional regularization, because they are scaleless when expanded in the ultrasoft scale. 

Since the NRY is organized as an expansion in $\alpha$ and $1/M$, we can readily understand which are the terms we need in the potential $V$ up to some order $M \alpha^n$ \cite{Pineda:1998kj,Pineda:1998kn}: one has to carry out the matching by combining inverse powers of the hard scale and the couplings as $(1/M)^a \alpha^b$ with $a+b \leq n-1$. In this work we aim at potential corrections up to order $M \alpha^4$. In practice, for a given diagram, one counts the powers of $\alpha$, inverse powers of $M$, and then multiplies by the soft scale $M \alpha$ in order to obtain the dimension of an energy. One has to consider tree-level, one- and two-loop diagrams for the matching similarly to what has been carried out in the pNRQED case \cite{Pineda:1998kn}. As explained above, in the diagrams involving heavy fermions in loops, we employ static fermion propagators, which allows us to simplify the derivation \cite{Manohar:1997qy,Pineda:1998kj}. We address here the tree-level diagram only, whereas a more detailed discussion on the loop diagrams is deferred to appendix \ref{app:loop_pot_match}.

\begin{figure}[t!]
    \centering
    \includegraphics[scale=0.61]{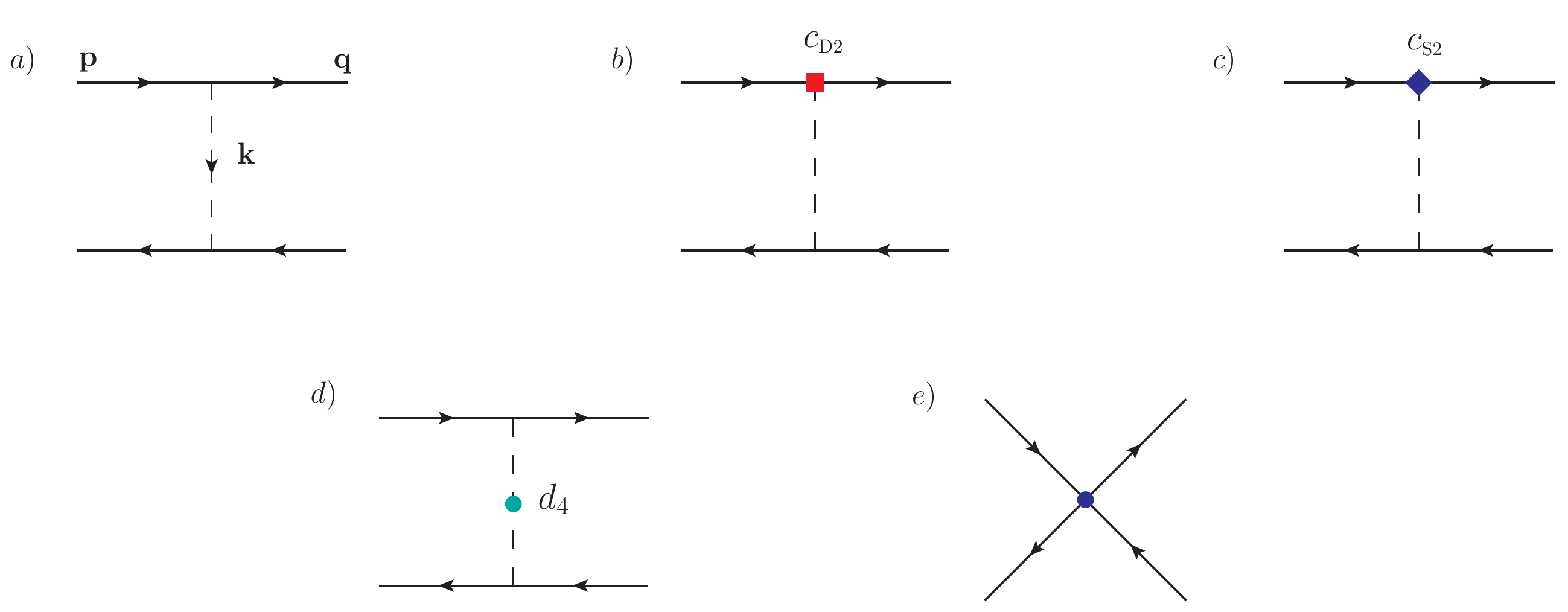}
    \caption{NREFT tree-level diagrams relevant for the potential matching. Upper row: The leftmost diagram is the leading contribution of order $M \alpha^2$, and we explicitly show the momentum labeling. The middle and rightmost diagrams involve $1/M^2$ suppressed vertices, and contribute at order $M \alpha^4$. Lower row: Additional diagrams at order $1/M^2$, which are, however, suppressed in the power counting $M \alpha^n$. The diagram on the left comprises the insertion of the effective scalar operator with the matching coefficient $d_4$, which starts at $\mathcal{O}(\alpha,\lambda)$. The right diagram stands for dimension-6 four-fermion interaction (it comprises both spin triplet and spin singlet).
    }
    \label{fig:match_pot_m2}
\end{figure}

We collect the relevant tree-level diagrams in figure \ref{fig:match_pot_m2}. In the upper row, the leftmost diagram provides the leading contribution to the potential of order $M \alpha^2$. It is easy to see that the middle and rightmost diagrams give instead a contribution of order $M \alpha^4$: the two inverse powers of $M$ in the vertex are compensated for the soft scale $M \alpha$, namely $\alpha \times 1/M^2 \times (M \alpha)^3$. A quantum correction to the matching coefficients $c_{\hbox{\tiny D}}$ and $c_{\hbox{\tiny S}}$ would give a contribution of order $M \alpha^5$. Next, let us discuss the diagrams in the lower row of figure \ref{fig:match_pot_m2}, as an example of diagrams which are  suppressed in the power counting. The leftmost diagram accounts for the insertion of the corrected scalar propagator, and the contribution scales as $ \alpha \times d_4 /M^2 \times (M \alpha)^3 \sim  d_4 M \alpha^4$. Since the matching coefficient is $d_4= \mathcal{O}(\alpha, \lambda)$, \ie it vanishes at tree-level, this diagram is beyond our desired accuracy. By applying similar power counting arguments, one sees that the 4-fermion dimension-6 operators would contribute at order $f(^1 S_0) M \alpha^3$ and $f(^1 S_0) M \alpha^3$. However, as shown in section \ref{sec:matching_dim_6_and_8}, such matching coefficients vanish at  $\mathcal{O}(\alpha)$ and, therefore, they could contribute only beyond the required accuracy. Finally, dimension-8 operators need not be considered, as they are further suppressed. We find that one- and two-loop diagrams do not contribute at order $M \alpha^4$, but only beyond (see discussion in appendix \ref{app:loop_pot_match}).
Then, the potential matching receives contributions only from the upper diagrams in figure \ref{fig:match_pot_m2}, and the corresponding ones with the vertices $c_{\hbox{\tiny D}}$ and $c_{\hbox{\tiny S}}$ on the antiparticle line (not displayed). The resulting potential reads 
\begin{equation}
    V(\bm{k}) = - \frac{ 4 \pi \alpha}{\bm{k}^2} + c_{\hbox{\tiny D}} \frac{ \pi \alpha }{M^2} \left( 1 - 4 \frac{\bm{p} \cdot \bm{k}}{\bm{k}^2} + 4 \frac{\bm{p}^2}{\bm{k}^2}\right)+i c_{\hbox{\tiny S}} \frac{2 \pi \alpha }{M^2} \frac{\bm{S} \cdot (\bm{p} \times \bm{k})}{\bm{k}^2},
    \label{pot_order_1overM2}
\end{equation}
where  $\bm{S}=\textrm{diag} (\bm{\sigma}_1,\bm{\sigma}_2)/2$ is the total spin matrix, with $\bm{\sigma}_1$ ($\bm{\sigma}_2$) being the spin matrix of the two-component particle (antiparticle) field in the fermion bilinear, and we used $c_{\hbox{\tiny D}}=-c'_{\hbox{\tiny D}}$ and $c_{\hbox{\tiny S}}=-c'_{\hbox{\tiny S}}$. As expected, the first term is the same as in pNRQED. This finding is consistent with the assumption $m \ll M \alpha$ and the contribution corresponds to a Coulomb potential upon performing the Fourier transform. The third term (up to a relative sign difference) equally appears in the pNRQED potential. Yet the second term is characteristic of the pNRY and differs from what one finds in pNRQED. This difference can be traced back to the different momentum combinations entering the non-relativistic Lagrangian in \Eq\eqref{NREFT_lag} as compared to the NRQED case. Upon performing the Fourier transform (\cf \eg \cite{adkins2013} for a collection of useful formulas), we obtain the following potential in the position space
 \begin{eqnarray}
    V(\bm{r})= - \frac{\alpha}{r} +c_{\hbox{\tiny D}}\frac{\pi \alpha}{M^2} \delta^3(\bm{r}) +c_{\hbox{\tiny D}} \frac{\alpha}{M^2 r} \bm{p}^2 - i c_{\hbox{\tiny D}} \alpha \frac{\bm{r} \cdot \bm{p}}{M^2 r^3} + c_{\hbox{\tiny S}}\frac{\alpha}{2 M^2 } \frac{\bm{L} \cdot \bm{S}}{r^3} \, ,
    \label{pot_order_1overM2_position}
\end{eqnarray}
with
\begin{equation}
    \bm{p} = -i \nabla_{\bm{r}}, \, \quad \bm{L} = \bm{r} \times \bm{p} \, .
\end{equation}
Using pNRY power counting rules ($p \sim M \alpha$, $r \sim 1/(M \alpha)$) one can directly read off the scaling of each term \Eq\eqref{pot_order_1overM2_position}: the Coulomb potential scales as $M \alpha^2$, while all the remaining terms are of order $M \alpha^4$. The term proportional to $ \bm{r} \cdot \bm{p}$ is a feature of the Yukawa fermion-scalar interaction and is not present in pNRQED.

We conclude this section by noticing that at leading order in $1/M$ and at $\mathcal{O}(r^0)$ in the multipole expansion, the equation of motion for $\varphi$ derived from the 
pNRY Lagrangian of \Eq\eqref{pNREFT_sca_0} takes the form of a Schr\"odinger equation 
\begin{equation}
    \left( i \partial_t + \frac{\bm{\nabla}^2}{M} -V^{(0)} -2 g \phi \right) \varphi =0 \,.
    \label{sch_pnyt}
\end{equation}
Solving this equation yields Coulombic energy levels $E_n$ and the Bohr radius $a_0$ given by
\begin{equation}
E_n = -\frac{M \alpha^2}{4 n^2} = -\frac{1}{M a_0^2 n^2}\, , \quad a_0 \equiv \frac{2}{M \alpha}\, .
    \label{Coulomb_en_levels}
\end{equation}
Let us also remark that since \Eq\eqref{sch_pnyt} describes a fermion-antifermion bound state,
it features a factor of 2 in front of the $-g\phi$ term and a potential $V^{(0)}$ as compared to \Eq\eqref{sch_nry} for a single fermion. 

Before moving to the applications of pNRY, one more comment is in order. It is well known that the Yukawa potential induced by a scalar mediator is universally attractive so that not only particle-antiparticle but also identical fermions can form bound states. This is very different from \eg QED, where $e^+ e^+$- or $e^- e^-$-interactions are repulsive. We have explicitly verified that the pNRY for particle-particle (antiparticle-antiparticle) pair admits the very same form as \Eq\eqref{pNREFT_sca_0} upon performing the replacements $\chi_j(t,\bm{x}_2) \to \psi^\dagger_j(t,\bm{x}_2)$ ($\psi^\dagger_i(t,\bm{x}_1) \to \chi_i(t,\bm{x}_1)$). However, since identical Dirac fermions cannot annihilate into scalars,
the bound-states $X X$ and $\bar{X} \bar{X}$ are completely stable in the context of the scalar Yukawa theory. 

 \subsubsection{Scalar mass of order \texorpdfstring{$M v$}{M v}}
 \begin{figure}
     \centering
     \includegraphics[scale=0.56]{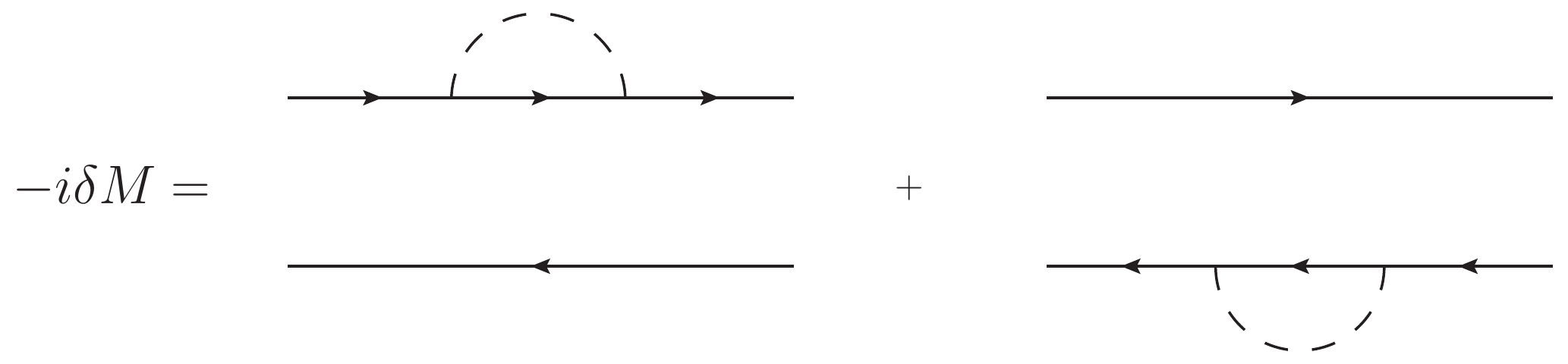}
     \caption{Matching condition for the mass contribution to the heavy pair propagator. The diagram topology has to be understood for the different tri-linear vertices in \Eq\eqref{NREFT_lag}. At the desired accuracy, contributing diagrams are those with two $c_1(c'_1)$ vertices, and one $c_1(c_1')$ and one $c_\textrm{D}(c'_\textrm{D})$ vertices. }
     \label{fig:finite_mass_shift}
 \end{figure}
The Yukawa potential is usually understood as a screened potential of the form $-\alpha e^{-mr}/r$. The mass of the mediator leads to a finite-range interaction with $r \sim 1/m$, as opposed to the Coulomb case. Our calculation of the potential in \Eq\eqref{pot_order_1overM2} assumed the scalar mass to be much smaller than the momentum transfer of order $Mv$ (we restore $v$ instead of $\alpha$ in order to be generic and not necessarily in the Coulomb regime $v \sim \alpha$). Hence, we consistently neglected the mediator mass in the matching and it did not appear in the corresponding potential in \Eq\eqref{pot_order_1overM2_position}.

Let us briefly discuss how the matching calculation changes when $m \sim Mv$, so that the scale $m$  cannot be neglected. The main difference resides in the scalar propagator that enters the upper row diagrams of figure~\ref{fig:match_pot_m2}, namely $-i/\bm{k}^2 \to -i /(\bm{k}^2 + m^2)$. The corresponding potential in the position space becomes
\begin{eqnarray}
 &&V_{m}(\bm{r})=  -\frac{\alpha e^{-mr}}{r}  +c_{\hbox{\tiny D}}\frac{ \pi \alpha}{M^2} \delta^3(\bm{r}) + c_{\hbox{\tiny D}}\frac{ \alpha}{M^2} \frac{e^{-mr}}{r} \left( \bm{p}^2-\frac{m^2}{4} \right)
 \nonumber
 \\
 && \hspace{1.6 cm} -i c_{\hbox{\tiny D}} \frac{\alpha e^{-m r}}{M^2 r^2} \left( m + \frac{1}{r}\right) \bm{r} \cdot \bm{p} + c_{\hbox{\tiny S}} \frac{\alpha e^{-m r}}{2 M^2 r^2} \left( m + \frac{1}{r}\right) \bm{L} \cdot \bm{S} \, ,
 \label{pot_order_1overM2_position_m_finite}
\end{eqnarray}
where one recognizes the leading term to be a Yukawa screened potential. Moreover, in this case the pole of the bound-state propagator receives a finite mass shift  \cite{Brambilla:2008cx,Ghiglieri:2013iya}.\footnote{Despite of the fact that these references discuss finite temperature calculations, the zero-temperature case follows the same pattern. The mediator mass $m$ is integrated out together with the inverse distance between the pair $1/r \sim Mv$. The potential and mass shifts are understood as matching coefficients of the pNREFT.} The corresponding one-loop diagrams are shown in figure~\ref{fig:finite_mass_shift}, and we find 
\begin{equation}
    \delta M = \alpha m - \frac{ c_{\hbox{\tiny D}} \alpha}{4 M^2} m^3 \, .
    \label{finite_delta_m}
\end{equation}
Notice that \Eq\eqref{pot_order_1overM2_position_m_finite} reduces to \Eq\eqref{pot_order_1overM2_position} in the limit $m \to 0$. On the other hand, one can also study corrections to the Coulombic regime by expanding \Eq\eqref{pot_order_1overM2_position_m_finite} in $m r \ll 1$.
Let us also remark that in the case of a vanishing scalar mass, the loop integrals in figure~\ref{fig:finite_mass_shift} are scaleless, and vanish accordingly in dimensional regularization. 

\subsection{Bound-state spectrum at order  \texorpdfstring{$M \alpha^4$}{M alpha4}}
\label{sec_pNREFT_application_1}
\begin{figure}
    \centering
    \includegraphics[scale=0.60]{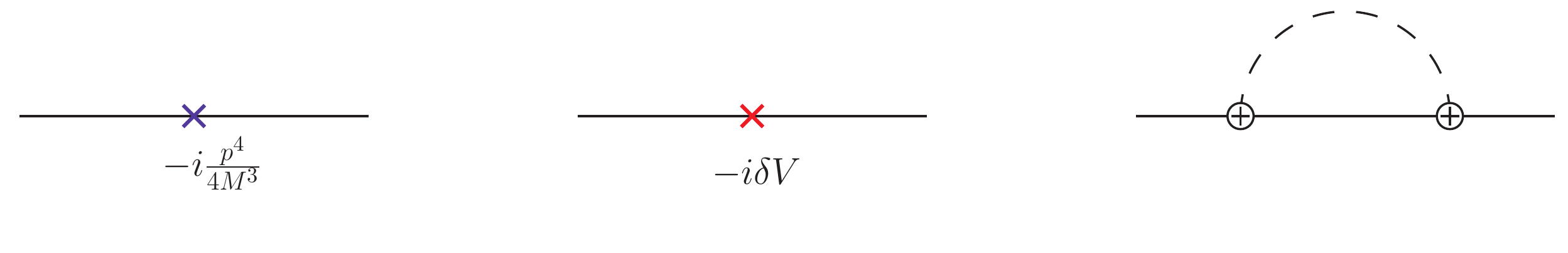}
    \caption{Diagrams needed for the derivation of the $M \alpha^4$ corrections to the bound-state spectrum. From left to right one finds: kinetic energy, correction from the potential and ultrasoft contribution from monopole interactions. The solid line represents a bound state $\varphi$, and $+$ depicts the monopole vertex. Quadrupole and derivative couplings, together with mixed combinations, are suppressed in the power counting.}
    \label{fig:delta_E}
\end{figure}
As a first non-trivial application of the pNRY, we carry out the derivation of the discrete spectrum at order $M \alpha^4$. We follow the setting of \cite{Pineda:1998kn,Pineda:1998kj} put forward for pNRQED and consider the two-point function of the field $\varphi$. The potential and kinetic contributions are simple insertions into the $\varphi$ propagator (see figure \ref{fig:delta_E} leftmost and middle diagrams). This brings us to the evaluation of quantum mechanical expectation values when projected onto bound-state wave functions. In the limit $m \ll M \alpha$, we can approximate the state as Coulombic and compute the expectation values on such unperturbed states. Additionally, one has to consider the ultrasoft contributions to the binding energy, namely those originating from the loop corrections to the propagator of an ultrasoft scalar. Similarly to the case of pNRQED, the leading non-vanishing ultrasoft contributions arise from one-loop self-energy diagrams. We assume that the scalar mass can be as large as the binding energy $M \alpha^2$. Applying the power counting of the pNRY, we see that only the diagram with two monopole vertices contributes within our accuracy, displayed in figure \ref{fig:delta_E}, and it scales as $M \alpha^3$. The ultrasoft contribution is finite, and some details of the calculation are provided in appendix \ref{app:loop_pot_match}. If the scalar mass is much smaller than the energy scale $M \alpha^2$, the scalar propagator can be expanded in $m /M\alpha^2$ and the corresponding loop integral vanishes in dimensional regularization. Accordingly, the one-loop self-energy diagram yields no contribution to the spectrum. 

Owing to the presence of the spin-orbit coupling in the Hamiltonian, the latter does not commute with $\bm{L}$ and $\bm{S}$ separately. However, the combination $\bm{L} \cdot \bm{S}$ can be rewritten in terms of the squared operators $\bm{J}^2$, $\bm{L}^2$ and $\bm{S}^2$ (and the Hamiltonian commutes with all of them), see \eg \cite{Sakurai:1167961}.
In this case it is common to label the states with $|n \ell j \rangle$, since $n$, $\ell$ and $j$ are good quantum numbers. The corrections to the spectrum  from the ultrasoft exchange $\delta E_{\hbox{\tiny US}}$,  kinetic energy $\delta E_{\textrm{kin}}$, and the potential terms $\delta E_{\delta V}$  read  
\begin{equation}
\delta E_{\hbox{\tiny US}} =  2 \alpha m \, , \quad    \delta E_{\textrm{kin}} =  \frac{M \alpha^4}{8} \frac{3\left( \ell + \frac{1}{2} \right)- 4 n}{4n^2(2 \ell +1)} \, ,
\end{equation}
and
\begin{eqnarray}
\delta E_{\delta V} &=& c_{\hbox{\tiny D}} \frac{M \alpha^4}{8 n^3} \delta_{s,1}\delta_{\ell,0} + c_{\hbox{\tiny D}} \left( -\frac{M \alpha^4}{8 n^4} + \frac{M \alpha^4}{2n^3 (2 \ell +1)} \right) + c_{\hbox{\tiny D}} \frac{M \alpha^4}{4 n^3} \delta_{s,1}\delta_{\ell,0} 
\nonumber
\\
&&+ c_{\hbox{\tiny S}} \frac{M \alpha^4}{8 n^3} \delta_{s,1}\delta_{\ell,0} + c_{\hbox{\tiny S}} \delta_{s,1}(1-\delta_{\ell,0}) \, \frac{M \alpha^4}{16 n^3 \ell (\ell +1) (2 \ell +1)} f_{j,\ell},
\end{eqnarray}
where 
\begin{equation}
     f_{j,\ell} = \left\{ \begin{array}{l}
       2 \ell,  \phantom{ccjjdddc} \text{for } j=\ell +1\\
        -2 , \phantom{ccddcdd} \text{for } j=\ell \\
        -2 (\ell +1) ,  \phantom{c} \text{for } j=\ell -1
        \end{array} \right. \, .
\end{equation}
A comment is in order on the form of the ultrasoft contribution $\delta E_{\hbox{\tiny US}}$. One can check its appearance in a complementary way. It has been already  stated
that this term corresponds to the propagation of an ultrasoft scalar with momentum/energy of order $M \alpha^2$. Assuming its mass to be of the same order, as done here, this amounts to expanding the potential in \Eq\eqref{pot_order_1overM2_position_m_finite} for $m r \ll 1$. The contribution at order $m \alpha$ from the potential expansion adds up with the one from $\delta M$ in \Eq\eqref{finite_delta_m}, giving $\delta E_{\hbox{\tiny US}}$, whereas the contribution at order $\alpha m^3$ cancels against the one in $\delta M$. In the case of a massive vector boson as a force mediator, the monopole contribution from the potential completely cancels against the mass correction (see \eg \cite{Brambilla:2008cx} for the QCD case) so that there is no analogue of $\delta E_{\hbox{\tiny US}}$ at order $m \alpha$.

\subsection{Bound-state formation cross section}
\label{sec_pNREFT_application_2}
Let us come to an application of the pNRY that establishes a connection to the recent developments in  DM phenomenology. As noted in the original works \cite{Feng:2009mn,vonHarling:2014kha}, the formation of unstable bound states can trigger another channel for DM annihilations, and consequently affect the estimates of the present-day relic density. In general, bound-state formation and decay are not only relevant for the early universe but can also provide enhanced signals in the annihilations of DM in the galactic halos and affect the corresponding experimental signatures. 

Here we deal with the bound-state formation via a radiative transition, where an above-threshold scattering state emits a scalar particle and turns into a bound state. In terms of the model degrees of freedom one has $(X\bar{X})_{\textrm{open}} \to (X\bar{X})_{\textrm{bound}}+ \phi$. This process occurs at the ultrasoft scale, 
so that the energy difference between the initial and the final state is of order $M \alpha^2$.
Such interactions can be naturally accommodated in our pNRY, where the field $\varphi$ accounts for both  scattering states (with positive energies) and  bound states  (with negative energies). Formally, one can think of it as of splitting $\varphi \equiv \varphi_s + \varphi_b$ in the particle-antiparticle Fock space. 
\begin{figure}
    \centering
    \includegraphics[scale=0.5]{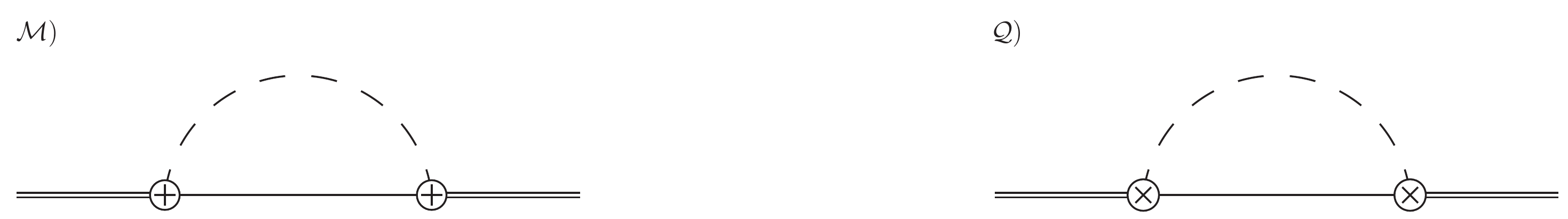}
    \caption{Self-energy diagrams in pNREFT for scattering states, whose imaginary part corresponds to the process $(X\bar{X})_{\textrm{open}} \to (X\bar{X})_{\textrm{bound}}+ \phi$. Solid double lines represent scattering states with positive energies, the monopole and quadrupole vertices are depicted as $+$ and $\times$ respectively. Diagrams with mixed monopole-quadrupole combinations are not shown. }
    \label{fig:MandQ_self}
\end{figure}

As it was done for the hard annihilations into scalars within NRY, here we again make use of the optical theorem. The process of interest can be calculated from the self-energy of the pair in a scattering state by  extracting its imaginary part. We show example diagrams in figure \ref{fig:MandQ_self}. Loop diagrams involve scales that are still dynamical in pNRY, namely the energy scale $M \alpha^2$ and the mass of the scalar $m$, which we have assumed to be much smaller than $M \alpha$ for the derivation of the pNREFT. No specific relation between their relative importance was needed in the matching between NRY \Eq\eqref{NREFT_lag} and pNRY \Eq\eqref{pNREFT_sca_0}. However, at this stage it is important to clarify their relative size. In the following derivation, we consider the case $m \siml M \alpha^2$, so that the $m$ must be retained in the scalar propagator. The method that we describe in the following is suitable for both in-vacuum and finite temperature calculations, provided that the thermal scales are smaller then the typical relative momentum of the pair. In this case, one can take the pNREFT Lagrangian from \Eq\eqref{pNREFT_sca_0} as a starting point  \cite{Escobedo:2008sy,Escobedo:2010tu,Brambilla:2008cx,Brambilla:2010vq} and incorporate $T \neq 0$ that would enter as a dynamical scale together with the in-vacuum parameters $m, M \alpha^2$.\footnote{The situation is of course different if the temperature and other thermal scales, for example thermal masses, are of comparable size or larger than $M \alpha$. Then the pNREFT has to be derived accordingly and it would be qualitatively different from \Eq\eqref{pNREFT_sca_0}. See \cite{Escobedo:2008sy,Escobedo:2010tu} for QED and \cite{Brambilla:2008cx,Brambilla:2010vq} for QCD.} We leave a comprehensive construction of EFTs for scalar mediators at finite temperature for future work on the subject. 

Let us come to the self-energy diagrams relevant for the derivation of the cross section. We show the ones that involve two monopole or two quadrupole vertices in figure \ref{fig:MandQ_self}. The calculation is done in dimensional regularization with $D=4 -2 \epsilon$. Let us start with the left diagram in figure \ref{fig:MandQ_self}, where the corresponding self-energy reads
\begin{equation}
     \Sigma_s^{\mathcal{M}} = -i 16 \pi \alpha \mu^{4-D} \, \int \frac{d^D k}{(2 \pi)^D} \frac{i}{P^0-h-k^0+i\eta} \frac{i }{(k^0)^2-E_\phi^2+i\eta} \,  \, ,
     \label{self_energy_mono}
 \end{equation}
with $E_\phi=\sqrt{\bm{k}^2+m^2}$ being the energy of the scalar mediator. In order for the self-energy to acquire a full meaning, it has to be projected on the external scattering states, labeled with the relative momentum quantum number $p=M v_{\textrm{rel}}/2$, so that $P^0=E_p=p^2/M$. Next, we also insert a complete set of bound states so that the internal propagator in the loop diagram describes indeed the propagation of the discrete states of the spectrum. More explicitly,
\begin{equation}
	\frac{i}{P^0-h-k^0+i\eta} = \sum_n \frac{i}{P^0-h-k^0+i\eta} \ket{n} \bra{n}  =
	\sum_n \frac{i}{E_p-E_n-k^0+i\eta}  \ket{n} \bra{n},
\end{equation}
with 
 \begin{equation}
     E_p-E_n \equiv \Delta E^p_n = \frac{p^2}{M} + \frac{M \alpha^2}{4 n^2}.
 \end{equation}
 One can easily extract the imaginary part of the self-energy using Cutkosky cutting rules at zero temperature \cite{Cutkosky:1960sp} that impose the kinematic condition $ 0 < k^0 < \Delta E^p_n $. The result for the inclusive cross section to produce all possible bound states reads
 \begin{equation}
     \sigma_{\textrm{bsf}}^{\mathcal{M}}v_{\textrm{rel}} =\langle \bm{p } | {2\rm{Im}}(-\Sigma^{\mathcal{M}}_s) | \bm{p} \rangle = 8 \alpha \, \sum_n \left[ (\Delta E^{p}_{n})^2 -m^2 \right]^{\frac{1}{2}} \, |\langle \bm{p} | n \rangle |^2 \, ,
     \label{cross_section_mono}
 \end{equation}
which has the correct dimension of inverse energy squared.\footnote{One can simply see this by recalling the energy dimension of bound and scattering states kets, given by $\left[ |n\rangle \right]=0$ and $\left[ |\bm{p}\rangle \right]=-3/2$ respectively.} 
 However, one can readily see that the cross section in \Eq\eqref{cross_section_mono} vanishes because of the orthogonality between the scattering and bound-state wave functions $\Psi_p(\bm{r})$ and $ \Psi_{n}(\bm{r}) $, that appear in the expectation value $
    \langle \bm{p} | n \rangle   = \int d^3 \bm{r} \,  \Psi^*_p(\bm{r}) \, \Psi_{n}(\bm{r}) =0$. For the same reason mixed monopole-quadrupole diagrams give no contribution to the total cross section. These findings nicely agree with the results of \cite{Wise:2014jva,An:2016kie}, where the authors consider interactions of Dirac fermion DM with a scalar mediator. The same pattern is observed also in the case of DM being a non-relativistic scalar particle coupled to a scalar mediator \cite{Petraki:2015hla,Petraki:2016cnz}. 
     
 Let us now consider the quadrupole contributions induced by the right diagram in figure \ref{fig:MandQ_self}. The corresponding self-energy reads
  \begin{eqnarray}
     \Sigma_s^{\mathcal{Q}} &=& -i \frac{\pi \alpha}{4} \mu^{4-D} r^i r^j \, \int \frac{d^D k}{(2 \pi)^D} \frac{i}{P^0-h-k^0+i\eta} \frac{i \, k^i k^j k^m k^n }{(k^0)^2-E_\phi^2 +i\eta} \, r^m r^n 
     \label{self_energy_quad}
 \end{eqnarray}
 where one may notice the appearance of powers of the scalar three momenta in  \Eq\eqref{self_energy_quad}, that are induced by the action of the derivative operator $\nabla_{\bm{R}}$ on the scalar propagator. This diagram can be evaluated in the same fashion as the  monopole contribution. The result for the cross section reads
 \begin{eqnarray}
     \sigma_{\textrm{bsf}}^{\mathcal{Q}}v_{\textrm{rel}} = \frac{\alpha}{120}    \sum_{n} \left[ (\Delta E_n^p)^2-m^2\right]^\frac{5}{2} \left[ |\langle\bm{p} | \bm{r}^2 |n  \rangle|^2 + 2|\langle\bm{p} | r^i r^j |n  \rangle|^2\right] \, ,
     \label{cross_section_quadrupole}
\end{eqnarray}
which also has the correct mass dimension and features non-trivial expectation values. The
form of the prefactor highlights the effect of the mediator mass being of the same order as the ultrasoft scale: this setting obviously features a strong suppression of the formation rate. On the other hand, in the case of $m \ll M \alpha^2$  one can simply expand the result in \Eq\eqref{cross_section_quadrupole} accordingly.

 \begin{figure}[t!]
        \centering
         \includegraphics[scale=0.61]{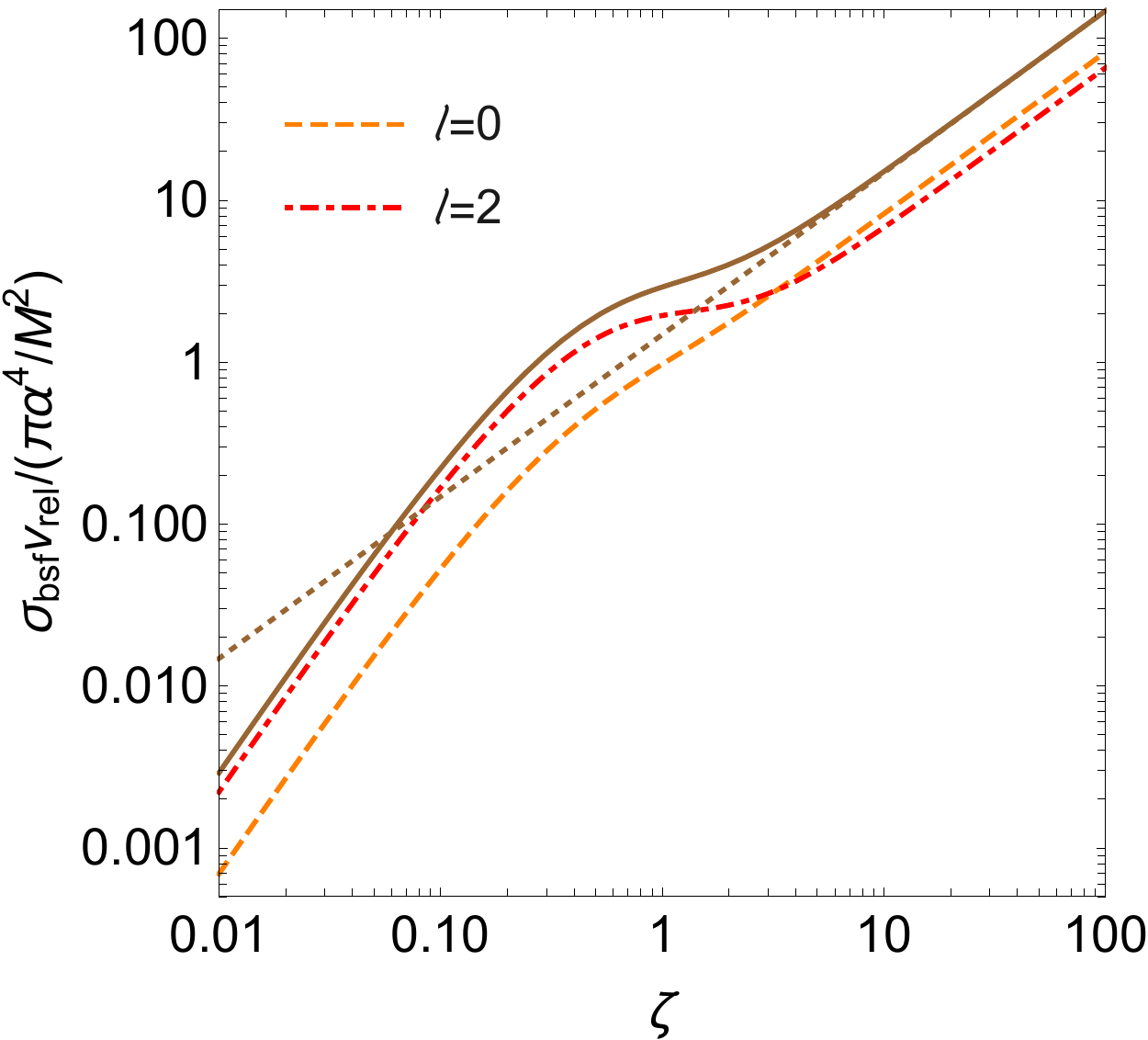}
        \caption{Bound-state formation cross section from \Eq\eqref{bsf_total} divided by $\pi \alpha^4/M^2$ as a function of $\zeta \equiv \alpha / v_\textrm{rel}$. 
        	The solid brown curve corresponds to the total cross section, while 
        	the dashed orange curve and dot-dashed red curve show the contributions from 
        	the $\ell=0$ and $\ell=2$ partial waves respectively. The behavior of the total cross section in the large $\zeta$ limit  \cite{An:2016kie} is represented by the dotted brown curve.
        }
        \label{fig:cross_bsf}
    \end{figure}

The total cross section also receives contributions from the spin-independent relativistic correction operator $- g \phi \bm{\nabla}^2_{\bm{r}}/M^2$ in \Eq\eqref{pNREFT_sca_0}. 
This amounts to additional diagrams involving two insertions of this operator, as well as diagrams with one such insertion and a quadrupole vertex, whereas the insertion of a monopole vertex yields a vanishing amplitude. Putting everything together, our final result for the total cross section reads
\begin{eqnarray}
 \sigma_{\textrm{bsf}} v_{\textrm{rel}} &=& \frac{\alpha}{120 }    \sum_{n} \left[ (\Delta E_n^p) ^2-m^2\right]^\frac{5}{2} \left[ |\langle\bm{p} | \bm{r}^2 |n  \rangle|^2 + 2 |\langle\bm{p} | r^i r^j |n  \rangle|^2\right] \nonumber
 \\
 && + 2 \alpha  \sum_{n} \left[ (\Delta E_n^p) ^2-m^2\right]^\frac{1}{2} \,  \Big \vert \Big \langle \bm{p} \Big \vert \frac{\nabla^2_{\bm{r}}}{M^2} \Big \vert n  \Big \rangle \Big \vert^{2} 
 \nonumber
 \\
 &&-  \frac{\alpha}{3}  \sum_{n} \left[ (\Delta E_n^p) ^2-m^2\right]^\frac{3}{2} \textrm{Re} \left[ \Big \langle \bm{p} \Big \vert \frac{\nabla^2_{\bm{r}}}{M^2} \Big \vert n  \Big \rangle  \langle n | \bm{r}^2 | p \rangle \right] \, .
 \label{bsf_matrix_element}
\end{eqnarray}
When considering the ground-state formation cross section, namely the state $|n\rangle=|100\rangle$, and neglecting the mediator mass, we obtain the following result using Coulomb bound state and scattering wave functions
 \begin{align}
       \sigma^{100}_{\textrm{bsf}} v_{\textrm{rel}} & = \frac{64}{9} \frac{\pi \alpha^4}{M^2}  \frac{ 2 \pi \zeta}{1-e^{-2 \pi \zeta}} \frac{\zeta^2}{1+\zeta^2} e^{-4 \zeta \arccot(\zeta)} 
       \nonumber
       \\
       & + \frac{256}{45}  \frac{\pi \alpha^4}{M^2} \frac{2 \pi \zeta}{1-e^{-2 \pi \zeta}} \frac{\zeta^2 (4+\zeta^2 )}{(1+\zeta^2)^2}e^{-4 \zeta \arccot(\zeta)} \, .
       \label{bsf_total}
    \end{align}
    Our result in \Eq\eqref{bsf_total} agrees with the earlier findings in the literature \cite{Wise:2014jva,An:2016kie} in the limit $\zeta \equiv \alpha / v_\textrm{rel} \gg 1$ . The first and the second term in \Eq\eqref{bsf_total} stem from the $\ell=0$ and $\ell=2$ partial waves in the scattering wave function respectively.\footnote{As done in ref.\cite{An:2016kie}, by choosing the coordinate system in such a way that $\bm{p}$ points into the $z$-direction, the scattering wave function can be expanded into partial waves as $\Psi_{p}(\bm{r}) = \sum_{\ell=0}^{\infty} \Psi_{p}^\ell(\bm{r})$ with $\Psi_{p}^\ell(\bm{r})=\langle\bm{r}|\bm{p}\ell\rangle$.}
    Their relative size as compared to the total cross section can be inferred from figure~\ref{fig:cross_bsf}, where they are depicted as dashed orange and dot-dashed red curves. The brown curves correspond to the total cross section, the solid line is our result \Eq\eqref{bsf_total}, whereas the dotted one is the large $\zeta$ limit \cite{Wise:2014jva,An:2016kie}.
    
    The recasting of the bound-state formation cross section in the language of pNRY offers a clear organization in terms of quantum mechanical expectation values. In particular, as for the ground-state formation, the $\ell=2$ contribution only comes from the $\langle\bm{p} | r^i r^j |n  \rangle$ matrix element (it develops a non-trivial angular dependence), whereas all the four matrix elements in \Eq\eqref{bsf_matrix_element} contribute to the $\ell=0$ term.

\section{Conclusions}
\label{sec_conclusions}
 Self-interacting dark matter is mostly welcome in the attempt of reproducing the observed galactic structures, and it appears to work better than collisionless dark matter. Typically, self-interactions between non-relativistic dark matter particles are induced by the exchange of a light mediator. In addition to the desired velocity-dependent interactions that accommodate the halos of different-sized objects,  
 it may well be that such self-interactions induce DM bound states. Most notably, depending on the model at hand in terms of its field content, masses and couplings, the impact of bound-state formation can play a rather important role in the determination of the present-day DM energy density. This may lead to sizable changes in the parameter space compatible with the cosmological abundance, making it necessary to revisit the relevant experimental bounds. 

In this work we studied a  model that represents a family of minimal DM models, where a light scalar mediator induces self-interactions between Dirac fermion DM via a Yukawa-type interaction. 
Making use of the assumed hierarchy of well separated dynamical scales
$M \gg Mv \gg M v^2$, we employed EFT techniques to study the resulting bound-state dynamics. In particular, we carried out a rigorous derivation of NRY and pNRY for a scalar force carrier, in the spirit of NRQED and pNRQED and their QCD counterparts. These EFTs are known to be very useful and successful tools for investigating and calculating observables
relevant to bound and near-threshold states. As for NRY we extended and generalized
the formulation already available in the literature \cite{Luke:1996hj,Luke:1997ys}, whereas, 
to the best of our knowledge, the explicit construction of pNRY was carried out in this paper for the first time.

We started with the derivation of the NRY from the first principles, where we identified the relevant degrees of freedom (non-relativistic Pauli fields and a scalar) and worked out the power-counting. We explicitly included $1/M^2$-operators in the bilinear sector and $1/M^4$-operators in the 
four-fermion sector, which allowed us to reproduce the first non-trivial contribution to the annihilation cross section $X \bar{X} \to \phi \phi$ at leading order in the fermion-scalar coupling. In the bilinear sector the matching was performed at tree-level.

Then, we resolved the power-counting ambiguity in the NRY due to the soft and ultrasoft scales being still intertwined, by constructing  the corresponding pNRY. The degrees of freedom of this low-energy theory were found to be particle-antiparticle pairs
(represented by a bilocal field) interacting with ultrasoft scalars. The scalars were enforced to be ultrasoft by performing a multipole expansion in the relative center of mass coordinate $r$. This way the presence of characteristic monopole and quadrupole interactions at the level of the pNRY Lagrangian was made manifest. The same is true also for the arising spin-independent relativistic correction with derivatives acting on the heavy-pair field. Dipole interactions that typically appear in pNRQED, turned out to be absent in pNRY.

We explicitly computed the  DM fermion-antifermion potential, which naturally arises
at the level of the pNRY Lagrangian as a matching coefficient. This paves way for a systematic inclusion of quantum and relativistic corrections
in future works on the topic. In the Coulombic $m \ll M \alpha$ regime of pNRY the scalar-induced potential turned out to share some similarities with the one in pNRQED.
However, we also found that the spin-orbit term comes with an overall opposite sing as compared to the pNRQED case, while the contribution induced by an operator proportional to
$\bm{r} \cdot (-i\nabla_{\bm{r}})$ has no correspondence in the electromagnetic potential.
We also performed the potential matching for the setting where the scalar mass is of order $M v$, thus recovering a Yukawa screened potential at leading order. Furthermore, we explained that pNRY can describe not only particle-antiparticle interactions but also bound states formed by identical Dirac fermions such as $XX$ and $\bar{X} \bar{X}$.

As a first application of the pNRY, we computed the bound-state spectrum at the next-to-leading order, namely $\mathcal{O}(M \alpha^4)$, which constitutes a new result presented in this work. In particular, we stressed the advantages of using a pNREFT for such bound state calculations as compared to non-EFT approaches. Our calculation was done in the Coulombic approximation, namely $m \ll M \alpha$, that still allows for the mediator mass to be as large as the ultrasoft scale $M \alpha^2$. The ultrasoft contribution to the spectrum, in the case of the scalar mass being not much smaller than the binding energy,  was found to provide the leading contribution of order $M \alpha^3$.

A further application of the pNRY that was presented in this paper is the derivation of  the bound-state formation cross section by taking the imaginary part of the heavy-pair field self-energy diagram. In particular, the contributions from monopole-induced diagrams turned out to be vanishing due to the orthogonality of the wave function of the discrete and continuous spectrum, in full agreement with the previous findings in the literature.
On the other hand, we identified the leading contribution to the cross section to be induced by quadrupole interactions and relativistic corrections. The final expression was written in terms of quantum mechanical expectation values that naturally arise in pNREFT calculations. 
We also performed an explicit analytic evaluation of these quantities for the Coulombic regime, thus agreeing with the earlier findings \cite{Wise:2014jva,An:2016kie} in the limit of $\zeta \gg 1$. The fact that we were able to obtain the previously unknown full analytic result for arbitrary $\zeta$-values using pNRY can be regarded as another highlight of this work.

To conclude, we would also like to provide a brief overview of future research directions
in this field in conjunction with NRY and pNRY. The minimal model addressed in our work can be varied in different ways, such as a Majorana DM rather than Dirac DM, or a more general interaction with a pseudo-scalar force carrier and cubic self-interactions of the scalar/pseudo-scalar mediator. These equally compelling realizations can be handled within the EFT approach presented here. Moreover, an accurate derivation of the relic density requires calculation of various processes (\eg bound-state formation, dissociation and Sommerfeld enhancement) to be done at finite temperature. We believe that the EFTs presented in this work can be regarded as a starting point for such finite temperature calculations, as it was the case with the corresponding generalizations of NRQED/NRQCD and pNRQED/pNRQCD. Especially in the heavy-ion phenomenology related to heavy quarkonia, pNRQCD has proven to be an extremely useful tool to scrutinize different hierarchies between in-vacuum and thermal scales, and to calculate relevant observables in a controlled and systematic way. It goes without saying that the presence of thermodynamical scales can significantly modify and affect the relevant cross sections also in the DM phenomenology. Therefore, the derivation of the finite temperature versions of NRY/pNRY constitutes a worthwhile and phenomenologically relevant task that we hope to address in the subsequent publications.


\section*{Acknowledgments}
The work of S.B. is supported by the Swiss National Science Foundation under the Ambizione grant PZ00P2\_185783. V.\,S. acknowledges the support from the DFG under grant 396021762 -- TRR 257 ``Particle Physics Phenomenology after the Higgs Discovery.'' The authors are grateful to Jacopo Ghiglieri for reading the manuscript and providing useful comments, and to Miguel Escobedo for stimulating discussions at early stages of our work. They would also like to thank Matthias Steinhauser for making them aware of \cite{Eiras:2006xm,Beneke:2015lwa} and Joan Soto for \cite{Pineda:2000sz}.
\appendix
\numberwithin{equation}{section}

\section{Matching coefficients of NRY}
\label{app:matching}
In this appendix, we provide a detailed derivation of the matching coefficients that enter the NRY Lagrangian \Eq\eqref{NREFT_lag}. As for the bilinear fermion (antifermion) sector we work at leading order and we discuss the derivation in section \ref{app:1_over_M2}. Then, in section \ref{sec:eom_1overM2} we derive the NRY by using the equation of motion method that allows us to (i) do a non-trivial check of the so-obtained matching coefficients at tree-level; (ii) write the NRY in a covariant fashion, implement the reparametrization invariance, here at order $1/M$, and consequently fix $c_2=1$ at all orders. Finally, in section \ref{sec:matching_dim_6_and_8} we provide the derivation of the matching coefficients of the dimension-6 and dimension-8 operators. 
\label{sec:matching}
\subsection{Matching of the fermion bilinear with scattering amplitude}
\label{app:1_over_M2}

The derivation of the matching coefficients for the fermion bilinear at order $1/M^2$ can be conducted in the following way. We write down the scattering amplitude for the process $\psi (p) \to \psi (q) + \phi(k) $ in the fundamental theory \Eq\eqref{lag_mod_0} and expand 
the resulting expression in powers of $p/M$, $q/M$. To this aim, we need to rewrite Dirac spinors in terms of the two-component Pauli spinors using non-relativistic normalization \cite{Bodwin:1994jh} 
\begin{equation}
    u(\bm{p}) = \sqrt{\frac{E_p+M}{2E_p}} \left( \begin{array}{c}
         \xi  \\
          \frac{\bm{p} \cdot \bm{\sigma}}{E_p+M} \xi
    \end{array} \right) \, , \quad  v(\bm{p}) = \sqrt{\frac{E_p+M}{2E_p}} \left( \begin{array}{c}
         \frac{\bm{p} \cdot \bm{\sigma}}{E_p+M} \eta   \\
          \eta
    \end{array} \right) \, ,
    \label{non_rel_spinor}
\end{equation}
with $E_p=\sqrt{p^2+M^2}$. Furthermore, we take the $\gamma$ matrices in the Dirac basis and decompose them into Pauli matrices. By momentum conservation at the vertex we have  $k=p-q$, which is the momentum carried by the scalar, so that the result for the diagram in figure \ref{fig:Fer_sca_match} (left) reads 
\begin{eqnarray}
   - i g \, \bar{u}(q) \phi(\bm{k}) u(p) &=&-i g \, \phi(\bm{k})  \sqrt{\frac{(E_p+M)(E_{q}+M)}{4 E_p E_{q}}} \left( \xi^\dagger , \xi^\dagger \frac{\bm{\sigma \cdot \bm{q}}}{E_{q}+M} \right) \gamma^0  \left( \begin{array}{c}
         \xi  \\
          \frac{\bm{p} \cdot \bm{\sigma}}{E_p+M} \xi
    \end{array} \right) \nonumber \\
    &=& -i g \, \phi(\bm{k}) \xi^\dagger \left( 1- \frac{(\bm{p}+\bm{q})^2}{8 M^2} - i \frac{(\bm{q}\times \bm{p})\cdot \bm{\sigma}}{4 M^2} \right) \xi + \mathcal{O}\left( \frac{1}{M^4} \right) \, , 
    \label{fermion_BL} 
\end{eqnarray}
for the particle interaction with $\phi$, whereas 
\begin{eqnarray}
   - i g \, \bar{v}(-q) \phi(\bm{k}) v(-p) &=&-i g \, \phi(\bm{k})  \sqrt{\frac{(E_p+M)(E_q+M)}{4 E_p E_q}} \left( - \eta^\dagger \frac{\bm{\sigma \cdot \bm{q}}}{E_{p}+M} , \eta^\dagger  \right) \gamma^0  \left( \begin{array}{c}
        -\frac{\bm{\sigma \cdot \bm{p}}}{E_{q}+M} \eta \\
         \eta
    \end{array} \right) \nonumber \\
    &=& i g \, \phi(\bm{k}) \eta^\dagger \left( 1- \frac{(\bm{p}+\bm{q})^2}{8 M^2} - i \frac{(\bm{q}\times \bm{p})\cdot \bm{\sigma}}{4 M^2} \right) \eta + \mathcal{O}\left( \frac{1}{M^4} \right) \, ,
    \label{anti_fermion_BL}
\end{eqnarray}
for the antiparticle. Upon identifying the 2-spinors $\xi$ and $\eta$ with the $\psi$ and $\chi$ fields respectively, we can compare the
so-obtained expressions to the amplitudes induced by NRY \Eq\eqref{NREFT_lag} and read off the values of the matching coefficients listed in \Eq\eqref{tree_level_bilinear}. It is interesting to remark that the Pauli structures appearing in \Eqs\eqref{fermion_BL} and \Eq\eqref{anti_fermion_BL} differ by an overall minus sign, which is different from the situation that one finds in NRQED (no sign difference). This can be traced back to the vector structure of the electromagnetic interaction that features an additional $\gamma$ matrix in the fermionic current, so that the couplings of the particle and the antiparticle to the temporal component of the photon field $A^0(k)$ have the same sign.

\subsection{Equations of motions method}
\label{sec:eom_1overM2}
This method exploits the equations of motion of the high- and low-energy excitations of the relativistic field $X$ \cite{Isgur:1989vq,Isgur:1989ed,Georgi:1990um,Eichten:1989zv,Manohar:2000dt} to derive the corresponding nonrelativistic EFT. The following derivation closely follows \cite{Manohar:2000dt}, where the same exercise is carried out for QCD.

 In practice, one starts from the full relativistic Lagrangian, where are we interested solely in the fermion-bilinear piece given by
\begin{equation}
    \mathcal{L}_{\textrm{heavy}}= \bar{X} (i\slashed{\partial}-g \phi -M)X \,.
\end{equation}
We then decompose the relativistic four-component spinor $X$ as 
\begin{equation}
    X=e^{-iM v \cdot x} (h_v (x) + H_v (x)) \, , \quad h_v=e^{iM v \cdot x} \frac{1+ \slashed{v}}{2} X \, \quad  H_v=e^{iM v \cdot x} \frac{1- \slashed{v}}{2} X \, ,
\end{equation}
where $(1\pm \slashed{v})/2$ are velocity-dependent projectors with $\slashed{v} \equiv v^\mu \gamma_\mu$. In the rest frame of the pair, where $v^\mu=(1,\bm{0})$, such operators reduce to  $(1\pm \gamma^0)/2$ and project onto the particle and antiparticle components of the Dirac field $X$. 

 Making use of the properties of the velocity projectors, we find
\begin{eqnarray}
    \mathcal{L}_{\textrm{heavy}}=\bar{h}_v \, ( i v \cdot \partial -g \phi ) \, h_v  - \bar{H}_v \, ( i v \cdot \partial + g \phi + 2 M ) \, H_v + \bar{h}_v i \slashed{\partial} H_v  + \bar{H}_v i \slashed{\partial} h_v \, ,
    \label{lag_eom_2}
\end{eqnarray}
where it is now clear that  $H_v$ comprises the large energy modes of order $M$ that we want to integrate out from the Lagrangian. 
To this aim, it is useful to introduce the derivative 
$\partial^\mu_{\perp}= \partial^\mu -  v \cdot \partial v^\mu$ and to rewrite the equation of motion for the field $H_v$ \begin{eqnarray}
    \frac{\delta \mathcal{L}}{\delta \bar{H}_v} =0 \, \Rightarrow \, H_v = \frac{1}{i v \cdot \partial + 2 M+ g \phi} i \slashed{\partial}_{\perp} h_v\, .
    \label{H_in_h}
    \end{eqnarray}
Substituting the expression for $H_v$ as given in (\ref{H_in_h}) into (\ref{lag_eom_2})  we find 
\begin{eqnarray}
    \mathcal{L}_{\textrm{heavy}}=\bar{h}_v \, ( i v \cdot \partial -g \phi ) \, h_v + \bar{h}_v i \slashed{\partial}_{\perp} \frac{1}{i v \cdot \partial + 2M + g \phi} i \slashed{\partial}_{\perp} h_v ,
\end{eqnarray}
which is still exact. Now we can directly expand $\mathcal{L}_{\textrm{heavy}}$ in $1/M$ and, up to order $1/M^2$, the Lagrangian reads
\begin{eqnarray}
    \mathcal{L}_{\textrm{heavy}}=\bar{h}_v \, \left( i v \cdot \partial -g \phi \right) h_v  - \bar{h}_v \frac{\slashed{\partial}_\perp^2  }{2M} h_v +\frac{g}{4M^2} \bar{h}_v \slashed{\partial}_\perp \phi \, \slashed{\partial}_\perp  h_v + \frac{g}{4M^2} \bar{h}_v \slashed{\partial}_\perp (i v \cdot \partial) \slashed{\partial}_\perp  h_v . 
    \nonumber
    \\
    \label{eom_lag_4}
\end{eqnarray}
In order to eliminate all terms containing $(v \cdot \partial)  h_v$ beyond $\mathcal{O}(1/M^0)$, we need to introduce a suitable field redefinition \cite{Manohar:1997qy} given by
\begin{equation}
h_v \to  h_v - \frac{\partial_\perp^2}{8M^2} h_v \,.
    \label{field_red_1overM2}
\end{equation}
Therefore, at $\mathcal{O}(1/M^2)$ we find
\begin{align}
    \mathcal{L}_{\textrm{heavy}} &= \bar{h}_v \, \left( i v \cdot \partial -g \phi \right) h_v  - \bar{h}_v \frac{\slashed{\partial}_\perp^2  }{2M} h_v +\frac{g}{8M^2} \bar{h}_v \left\lbrace  \phi , \partial_\perp^2  \right\rbrace h_v \nonumber \\
    & + \frac{g}{4M^2} \bar{h}_v \slashed{\partial}_\perp \, \phi \, \slashed{\partial}_\perp  h_v . 
    \label{eom_lag_5}
\end{align}

Let us stress that the Lagrangian in \Eq\eqref{eom_lag_5} may describe not only non-relativistic systems made of heavy Dirac fermions of the same mass, but also bound states formed out of a heavy and a light fermion, which might be another interesting DM scenario worth exploring in more details using our EFT framework. This statement is completely analogous to the well-known fact \cite{Manohar:1997qy} that the HQET Lagrangian is equally suitable for studying properties of heavy-light mesons and heavy quarkonia: Both theories share the same Lagrangian but differ in their power-counting.

To complete our derivation for the case of the NRY, we switch to the rest frame with $v^\mu =(1,\bm{0})$, employ the relation $\sigma^i \sigma^j=\delta^{ij}+ i \epsilon^{ijk} \sigma^k$ and identify $h_v$ with the particle component $\psi$ of the $X$ field. Thus, we obtain
\begin{eqnarray}
    \mathcal{L}_{\textrm{heavy}}=\psi^\dagger \, \left( i \partial_0  -g \phi   + \frac{\bm{\nabla}^2}{2 M} -g \frac{i  \sigma^k \epsilon^{ijk} \bm{\nabla}^i\phi \bm{\nabla}^j}{4 M^2}-g \frac{ \bm{\nabla}^i\phi \bm{\nabla}^i}{4 M^2}-g \frac{ \left\lbrace \bm{\nabla} ^2  , \phi \right\rbrace }{8 M^2}\right) \psi \, ,
    \label{eom_lag_6}
\end{eqnarray}
that agrees with the particle sector of \Eq\eqref{NREFT_lag} with $c_1=c_{\hbox{\tiny D}}= c_{\hbox{\tiny S}}=-1$, $c_2=1$  and $c_3=c_4=0$. The piece of the NRY Lagrangian describing the antiparticle can be obtained in a similar way. 

\subsection{Foldy-Wouthuysen-Tani method}

The main idea behind the Foldy–Wouthuysen-Tani (FWT) \cite{Foldy:1949wa,Tani:1951} method is to introduce a sequence of unitary transformations that decouple the upper and lower components of the Dirac spinor order by order in $1/M$. Consequently, in the non-relativistic limit the Dirac equation splits into two separate equations for Pauli fields describing particles and antiparticles respectively. The procedure of applying FWT transformations to QED can be found in various QFT textbooks (\cf \eg \cite{Bjorken:1965sts,Itzykson:1980rh,Rebhan:2010} that we will partially follow here) and is often taught in advanced quantum mechanics courses. Therefore, we do not claim any originality for most of the material presented below. Once the technicalities behind the QED case are understood, it is a simple exercise to repeat the same procedure for the scalar Yukawa theory. The results for the pseudoscalar case can be found in \cite{Platzman:1960dqa}.

First of all, let us introduce the concept of  \emph{even} and \emph{odd} operators. Even operators are those that do not interchange upper and lower components of the Dirac spinor $X$, so that particles and antiparticles remain decoupled. Odd operators, on the contrary, are responsible for the mixing between particles and antiparticles. Schematically, we can write
\begin{equation}
	\hat{E} \begin{pmatrix}
		\psi \\ \chi
	\end{pmatrix} =  \begin{pmatrix}
		\# \psi \\ \# \chi
	\end{pmatrix}, \quad \quad \hat{O} \begin{pmatrix}
	\psi \\ \chi
\end{pmatrix} = \begin{pmatrix}
\# \chi \\ \# \psi
\end{pmatrix},
\end{equation}
where $\hat{E}$ is an even and $\hat{O}$ is an odd operator. In the context of the Dirac Hamiltonian we have 
$\alpha^i = \gamma^0 \gamma^i$ and $\beta = \gamma^0$, where the former is odd, while the latter is even. The Dirac spinor field $X$ satisfies 
\begin{equation}
	i \partial_t X = \hat{H} X,
\end{equation}
which yields the familiar Dirac equation in the case of a non-interacting Hamiltonian. 

For the sake of clarity, let us first discuss the generic case, without making an explicit reference to a particular theory. Our starting point for applying the FWT procedure is the unitary transformation
\begin{equation}
	X \to X' = \hat{U} X,
\end{equation}
with $\hat{U} = e^{i \hat{S}}$, where $\hat{S}$ is some operator. Then the time evolution of the transformed field becomes
\begin{equation}
	i \partial_t X' = e^{i \hat{S}} ( \hat{H} - i \tilde{\partial}_t ) e^{-i \hat{S}} X'.
\end{equation}
Here $\tilde{\partial}_t$ means that the partial derivative acts only on $e^{-i \hat{S}}$ but not on $X'$. Therefore, the transformed field satisfies
\begin{equation}
	i \partial_t X' = \hat{H}' X',
\end{equation}
where
\begin{equation}
	\hat{H}' = \hat{U} (\hat{H} - i \tilde{\partial}_t) \hat{U}^\dagger.
\end{equation}
Using the Baker–Campbell–Hausdorff formula we can rewrite $\hat{H}'$ as 
\begin{align}
	\hat{H}' &= \hat{H} + [i \hat{S},\hat{H}] + \frac{1}{2!} [i\hat{S},[i\hat{S},\hat{H}] ] + \frac{1}{3!} [i\hat{S},[i\hat{S},[i\hat{S},\hat{H}] ]] + \ldots \nonumber \\
	&  - \dot{\hat{S}} - \frac{1}{2!}  [i \hat{S}, \dot{\hat{S}} ] - \frac{1}{3!} [i \hat{S},  [i \hat{S}, \dot{\hat{S}} ] ]  + \ldots,
\end{align}
where we used that $\tilde{\partial}_t$ is non-vanishing only when it acts on a time-dependent function.

In the case of an interacting theory (e.\,g. QED) it is usually not possible to choose an $\hat{S}$ such, that 
the upper and lower components of $X$ decouple to all order in the $1/M$ expansion. Instead, one proceeds by starting with an ansatz that removes all odd terms at $\mathcal{O}(1/M^0)$ and then calculates $\hat{H}'$ to the desired order in $1/M$, say $1/M^2$. The resulting Hamiltonian contains odd terms at  $\mathcal{O}(1/M)$ which can be removed by applying a new unitary transformation $\hat{U}' = e^{i \hat{S}'}$, that requires us to evaluate
\begin{align}
	\hat{H}'' &= \hat{H}' + [i \hat{S}',\hat{H}'] + \frac{1}{2!} [i\hat{S}',[i\hat{S}',\hat{H}'] ] + \frac{1}{3!} [i\hat{S}',[i\hat{S}',[i\hat{S}',\hat{H}'] ]] + \ldots \nonumber \\
	&  - \dot{\hat{S}}' - \frac{1}{2!}  [i \hat{S}', \dot{\hat{S}}' ] - \frac{1}{3!} [i \hat{S}',  [i \hat{S}', \dot{\hat{S}}' ] ]  + \ldots.
\end{align}
This procedure needs to be iterated order by order in $1/M$ until all odd terms at the desired order in $1/M$ have been removed.

To find an ansatz for $\hat{S}$ it is useful to rewrite the initial Hamiltonian as
\begin{equation}
	\hat{H} = M \beta + \hat{\mathcal{E}} + \hat{\mathcal{O}},
\end{equation}
where $M \beta$ denotes the mass term, while $\hat{\mathcal{E}}$ and $\hat{\mathcal{O}}$ stand for the even and odd terms respectively. Notice that both $\hat{\mathcal{E}}$ and $\hat{\mathcal{O}}$ are of $\mathcal{O}(1/M^0)$. Let us now consider
\begin{align}
	\hat{H}' & = \hat{H} + [i \hat{S},\hat{H}] + \dot{\hat{S}} + \ldots = M \beta + \hat{\mathcal{E}} + \hat{\mathcal{O}} + [i \hat{S}, M \beta]  + \underbrace{[i \hat{S},\hat{\mathcal{E}}]}_{\equiv \hat{\mathcal{E}}' } + \underbrace{[i \hat{S},\hat{\mathcal{O}}]}_{\equiv \hat{\mathcal{O}}' } + \dot{\hat{S}} + \ldots
\end{align}
Since $\hat{S}$ and $\dot{\hat{S}}$ are $1/M$ suppressed as compared to $\hat{H}$, the same holds also for $\hat{\mathcal{E}}'$ and $\hat{\mathcal{O}}'$. Therefore, in order to remove the odd-term $\hat{\mathcal{O}}$ at  $\mathcal{O}(M^0)$ we need to choose an $\hat{S}$
that satisfies
\begin{equation}
	\hat{\mathcal{O}} + [i \hat{S}, M \beta ] = 0.
	\label{eq:ansatz-cond}
\end{equation}
If the odd piece is a linear combination of terms that anticommute with $\beta$ (e.\,g. terms proportional to $\alpha^i$ or $\beta \gamma^5$), we have
\begin{equation}
	\hat{\mathcal{O}} \beta = - \beta \hat{\mathcal{O}},
\end{equation}
which implies that 
\begin{equation}
	\hat{S} = c \beta \hat{\mathcal{O}},
\end{equation}
with $c$ being a normalization factor, is a suitable ansatz. Plugging this into \Eq\eqref{eq:ansatz-cond} we find
\begin{equation}
	\hat{O} + c [\beta \hat{\mathcal{O}}, M \beta ] = \hat{\mathcal{O}} - 2 i c M \hat{\mathcal{O}} \overset{!}{=0},
\end{equation}
which yields
\begin{equation}
	i \hat{ S} = \frac{1}{2 M} \beta \hat{\mathcal{O}}.
\end{equation}
Notice that we do not need to remove $M \beta$ using $\hat{S}$. That term can be taken care of later by a special unitary transformation of the Pauli spinors
\begin{equation}
	\psi \to e^{-i M t} \psi, \quad \chi \to e^{i M t} \chi \label{eq:massUT}.
\end{equation}
The above procedure of determining the proper ansatz can be also iterated at higher orders. For example, consider $\hat{H}'$, where the odd-terms may appear only at $\mathcal{O}(1/M)$ and higher. When constructing $\hat{H}''$ we need to keep in mind that our new $\hat{S}'$ is $1/M^2$ suppressed as compared to the original $\hat{H}$. Hence,
\begin{equation}
	\hat{H}'' = \hat{H}' + [i \hat{S}',\hat{H}'] + \ldots = M \beta + \hat{\mathcal{E}}' + \hat{\mathcal{O}}' + [i \hat{S}', M \beta] + \underbrace{[i \hat{S}',\hat{\mathcal{E}}']}_{\equiv \hat{\mathcal{E}}'' } + \underbrace{[i \hat{S},\hat{\mathcal{O}}']}_{\equiv \hat{\mathcal{O}}'' }  + \ldots
\end{equation}
which leads us to the following requirement at $\mathcal{O}(1/M)$
\begin{equation}
	\hat{\mathcal{O}}' + [i \hat{S}', M \beta ] = 0 \, .
\end{equation}
If $\hat{\mathcal{O}}'$ contains an odd number of $\alpha^i$ matrices in each term, then the relation
\begin{equation}
	\{ \hat{\mathcal{O}}', \beta \} = 0 \, ,
\end{equation}
clearly holds and we may continue as in the case of $\hat{S}$. That is,
\begin{equation}
	i \hat{S}' = \frac{1}{2 M} \beta \hat{\mathcal{O}}'.
\end{equation}
Otherwise one would have to choose a different ansatz for $\hat{S}'$. In practice, we may try
to employ the relation
\begin{equation}
	i \hat{S}^{n'} = \frac{1}{2 M} \beta \hat{\mathcal{O}}^{n'} \, ,
\end{equation}
and then check whether this ansatz indeed removes all odd terms at the given order in $1/M$.

Let us now specialize to the scalar Yukawa theory, where
\begin{equation}
	\hat{H} = \alpha \cdot \bm{\hat{p}} + M \beta + g \beta \phi
\end{equation}
and
\begin{equation}
	\hat{S} = \frac{-i}{2 M} \beta \alpha \cdot \bm{\hat{p}}.
\end{equation}
Using the familiar relations between $\alpha$ and $\beta$ matrices
\begin{equation}
	\{ \alpha^i, \alpha^j \} = 2 \delta^{ij} = - 2 \eta^{ij}, \quad \{ \alpha^i, \beta \} = 0, \quad \beta^2 = 1,
\end{equation}
it is easy to show that
\begin{equation}
	\quad (\alpha \cdot \bm{\hat{p}})^2 = \bm{\hat{p}}^2, \quad
	[\beta \alpha \cdot \bm{\hat{p}}, \alpha \cdot \bm{\hat{p}}] = 2 \beta \bm{\hat{p}}^2, \quad 
	[\beta \alpha \cdot \bm{\hat{p}}, \beta] = -  2 \alpha \cdot \bm{\hat{p}}.
\end{equation}
Consequently, up to $\mathcal{O}(1/M^2)$  we find
\begin{align}
	\dot{\hat{S}} &= 0, \\
	[i \hat{S},\hat{H}] &= \beta \frac{\bm{\hat{p}}^2}{M} - \alpha \cdot \bm{\hat{p}}  - \frac{g}{2 M} \alpha^i \{\phi, \bm{\hat{p}}^i\}, \\
	[i\hat{S},[i\hat{S},\hat{H}] ] &= -\frac{\bm{\hat{p}}^2}{M^2} \alpha \cdot \bm{\hat{p}} - \beta \frac{\bm{\hat{p}}^2}{M} - \frac{g}{4 M^2} \beta \alpha^i  \alpha^j \left ( \bm{\hat{p}}^i \{ \bm{\hat{p}}^j, \phi \} + \{\phi, \bm{\hat{p}}^i  \} \bm{\hat{p}}^j  \right ),\\
	[i\hat{S},[i\hat{S},[i\hat{S},\hat{H}] ]] &= -\frac{\bm{\hat{p}}^2}{M^2} \alpha \cdot \bm{\hat{p}},
\end{align}
so that
\begin{align}
	\hat{H}' &=  \beta \frac{\bm{\hat{p}}^2}{2 M} - \frac{g}{2M} \alpha^i \{ \phi, \bm{\hat{p}}^i\} - \frac{\bm{\hat{p}}^2}{3 M^2} \alpha \cdot \bm{\hat{p}} - \frac{g}{8 M^2} \beta \alpha^i  \alpha^j \left ( \bm{\hat{p}}^i \{ \bm{\hat{p}}^j, \phi \} + \{\phi, \bm{\hat{p}}^i  \} \bm{\hat{p}}^j  \right )  \nonumber \\
	& +  M \beta + g \beta \phi.
\end{align}	
Notice that odd operators present in  $\hat{H}'$ are of $\mathcal{O}(1/M)$. To remove the term proportional to $\alpha^i \{ \phi, \bm{\hat{p}}^i\}$ at $\mathcal{O}(1/M)$, we need to introduce a new unitary transformation
\begin{equation}
\hat{S}' = \frac{i g}{4 M^2} \beta \alpha^i \{ \phi, \bm{\hat{p}}^i\},
\end{equation}
with
\begin{equation}
	\dot{\hat{S}}' = \frac{i g}{4 M^2} \beta \alpha^i \{ \partial_0 \phi, \bm{\hat{p}}^i\}.
\end{equation}
The only non-vanishing commutator at $\mathcal{O}(1/M^2)$ is given by
\begin{align}
	[i \hat{S'},\hat{H'}] &=  \frac{g}{2M} \alpha^i  \{ \phi, \bm{\hat{p}}^i \} +  \frac{\bm{\hat{p}}^2}{3 M^2} \alpha \cdot \bm{\hat{p}} + \frac{g^2}{4 M^2} \alpha^i \{ \phi, \{ \phi, \bm{\hat{p}}^i \} \}.
\end{align}
This yields
\begin{align}
	\hat{H}'' &=  \beta \frac{\bm{\hat{p}}^2}{2 M}   - \frac{g}{8 M^2} \beta \alpha^i  \alpha^j \left ( \bm{\hat{p}}^i \{ \bm{\hat{p}}^j, \phi \} + \{\phi, \bm{\hat{p}}^i  \} \bm{\hat{p}}^j  \right )  \nonumber \\
	& + \frac{g^2}{4 M^2} \alpha^i \{ \phi, \{ \phi, \bm{\hat{p}}^i \} \} - \frac{i g}{4 M^2} \beta \alpha^i \{ \partial_0 \phi, \bm{\hat{p}}^i\}  +  M \beta + g \beta \phi.
\end{align}
To eliminate the two odd terms at $\mathcal{O}(1/M^2)$ we introduce
\begin{equation}
	\hat{S}'' = -\frac{i g^2}{8 M^3} \beta  \alpha^i \{ \phi, \{ \phi, \bm{\hat{p}}^i \} \} - \frac{g}{8 M^3} \alpha^i \{ \partial_0 \phi, \bm{\hat{p}}^i\},
\end{equation}
and upon picking up the $\mathcal{O}(1/M^2)$ relevant contributions from $\dot{\hat{S}}''$ and
\begin{equation}
	[i \hat{S''},\hat{H''}] =  - \frac{g^2}{4 M^2} \alpha^i \{ \phi, \{ \phi, \bm{\hat{p}}^i \} \} + \frac{i g}{4 M^2} \beta \alpha^i \{ \partial_0 \phi, \bm{\hat{p}}^i\},
\end{equation}
we arrive at our final  $\mathcal{O}(1/M^2)$ Hamiltonian that is free of odd operators
\begin{equation}
	\hat{H}''' =  \beta \frac{\bm{\hat{p}}^2}{2 M}  - \frac{g}{8 M^2} \beta \alpha^i  \alpha^j \left ( \bm{\hat{p}}^i \{ \bm{\hat{p}}^j, \phi \} + \{\phi, \bm{\hat{p}}^i  \} \bm{\hat{p}}^j  \right )    +  M \beta + g \beta \phi.
\end{equation}
Going back to the Lagrangian form, we find
\begin{equation}
	\mathcal{L} = X^{'''\dagger} (i \partial_t - \hat{H}''' ) X'''.
\end{equation}
Using
\begin{equation}
	\beta \begin{pmatrix}
		\psi \\ \chi
	\end{pmatrix} = \begin{pmatrix}
		\psi \\ - \chi
	\end{pmatrix}, \quad 	\alpha^i \begin{pmatrix}
		\psi \\ \chi
	\end{pmatrix} = \begin{pmatrix}
		\sigma^i \chi \\ \sigma^i \psi 
	\end{pmatrix}
\end{equation}
and applying the unitary transformations given in \Eq\eqref{eq:massUT} one readily obtains the fermion-bilinear part of the Lagrangian given in \Eq\eqref{NREFT_lag}, excluding operators with vanishing tree-level matching coefficients.

\subsection{Matching of the dimension-6 and dimension-8 operators}
\label{sec:matching_dim_6_and_8}
Here we closely follow the tree-level matching between QCD and NRQCD in the 4-fermion sector described in \cite{Bodwin:1994jh}. We derive the contribution to the amplitude $X\bar{X} \to  X\bar{X}$  in the center of mass reference frame. Hence, we take the incoming $X$ and $\bar{X}$ to have momenta $\bm{p}$ and $-\bm{p}$, whereas the outgoing $X$ and $\bar{X}$ have momenta $\bm{p}'$ and $-\bm{p}'$ respectively. Due to energy conservation and same masses involved in the process for the DM states, one has  $|\bm{p}|=|\bm{p}'| \equiv p$.\footnote{The NREFTs can be also formulated for particles with different masses, say $M_1$ and $M_2$, see \cite{Pineda:1997bj,Pineda:1998kn}. Here we consider the equal mass case only.} The form of the non-relativistic Dirac spinors has been already   given in \Eq\eqref{non_rel_spinor}. The matching is performed by enforcing on-shell four-fermion Green's function in the full theory (\ref{lag_mod_0}) and in the NRY (\ref{NREFT_lag}). 
\begin{figure}[t!]
    \centering
    \includegraphics[scale=0.55]{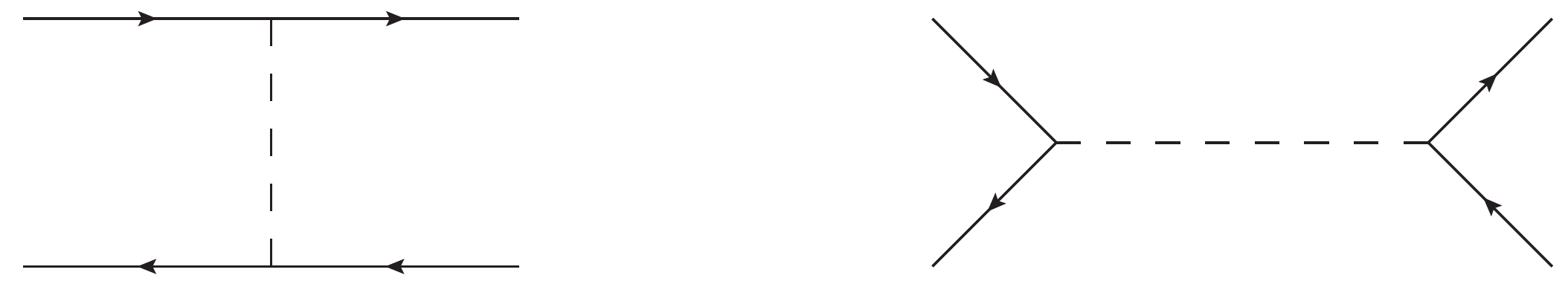}
    \caption{Leading order 4-fermion diagrams. The diagram on the left also appears in the NRY, so that it does not enter the matching.}
    \label{fig:sca_alpha}
\end{figure}

For completeness, let us briefly describe the matching of the four-fermion operators at order $\alpha$. The relevant tree-level diagrams are shown in figure \ref{fig:sca_alpha}. It is clear that no imaginary part can arise at this order. Moreover, the diagram on the left
can be precisely reproduced in the NRY because the scalar can carry a soft momentum (it is indeed the diagram appearing in the potential matching in figure \ref{fig:match_pot_m2}). Only the diagram on the right contributes to the matching at this order, and provides a contribution to the real part of the matching coefficients. We find the only non-vanishing coefficient at order $\alpha$ to be  $f(^3P_0) = 3 \pi \alpha$, while the matching coefficients of all dimension-6 operators vanish.
Going to order $\alpha^2$, we find two one-loop diagrams that contribute to the process $X \bar{X} \to \phi \phi$ and we show them in figure \ref{fig:DM_0_ann_sca}. We are interested in their imaginary parts, which can be extracted by using the standard cutting rules, namely putting the internal scalar fields on shell. Expanding up to second order in the velocities $\bm{v}=\bm{p}/E$, $\bm{v}'=\bm{p}'/E$, and writing the final result as in \cite{Bodwin:1994jh}, we find the annihilation contribution to the scattering amplitude to be
\begin{eqnarray}
    {\rm{Im}}(-i\mathcal{M}_{t+u}) = \frac{\pi \alpha^2}{2 M^2} \left[ \frac{1}{15} \bm{v}' \cdot \bm{v} \, \sigma^i \otimes \sigma^i +  \frac{41}{15}  \bm{v}' \cdot \bm{\sigma} \otimes \bm{v} \cdot \bm{\sigma}  + \frac{1}{15} \bm{v} \cdot \bm{\sigma} \otimes \bm{v}' \cdot \bm{\sigma}  \right] \, ,
    \label{NREFT_t_and_u}
\end{eqnarray}
where the subscript $t+u$ indicates the sum of the $t$- and $u$-channel diagrams of figure \ref{fig:DM_0_ann_sca}. 
We remark that there is no term of order $v^0$, implying that in the Yukawa theory (\ref{lag_mod_0}) annihilations are velocity suppressed and start at order $v^2$. The matching coefficients of the dimension-6 operators are zero, also at order $\alpha^2$ (as expected by symmetry arguments). On the contrary, some of the Lagrangian terms in \Eq\eqref{dimension_8_lag} contribute to the matching, and one obtains
\begin{eqnarray}
\frac{{\rm{Im}}[f(^3P_1)]+{\rm{Im}}[f(^3P_2)]}{2} &=&  \frac{1}{30} \pi \alpha^2   \, , \\
 \quad \frac{{\rm{Im}}[f(^3P_0)]-{\rm{Im}}[f(^3P_2)]}{3} &=&  \frac{41}{30} \pi \alpha^2  \, , 
 \\
 \frac{{\rm{Im}}[f(^3P_2)]-{\rm{Im}}[f(^3P_1)]}{2}   &=& \frac{1}{30} \pi \alpha^2   \, ,
\end{eqnarray}
that brings us to the result given in \Eq\eqref{match_sca_dim_8}.

\section{Loop contributions}

\subsection{Loop diagrams for potential matching of pNRY}
\label{app:loop_pot_match}
\begin{figure}[t!]
    \centering
    \includegraphics[scale=0.55]{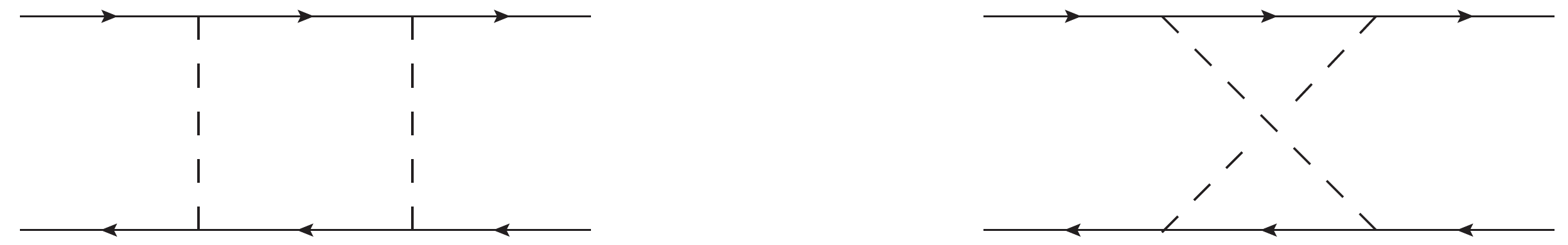}
    \caption{One-loop diagrams of order $M \alpha^3$, the sum of which cancel.}
    \label{fig:match_one_loop_sca_1}
\end{figure}

In this section we would like to discuss one- and two-loop diagrams that need to be analyzed for the potential matching of the pNRY. The systematic analysis is partly based on the pNRQED matching in the Feynman gauge \cite{Pineda:1998kn}, where the temporal component of the photon field has to be considered in loop diagrams (at variance with the Coulomb gauge). Then we have to consider (i) one loop diagrams as given in figure \ref{fig:match_one_loop_sca_1}  (possible contribution at $\mathcal{O}(M \alpha^3)$); (ii) the same diagrams with a kinetic insertion $\bm{p}^2/2M$ in one of the fermion lines at a time, (possible contribution at $\mathcal{O}(M \alpha^4)$); (iii) again the same diagrams with external energy insertions arising from the expansion of the propagators around zero external energy (possible contribution at $\mathcal{O}(M \alpha^4)$); (iv) two-loop diagrams involving scalar propagators without kinetic/external energy insertion (possible contribution at $\mathcal{O}(M \alpha^4)$). We checked explicitly that the same arguments put forward for the QED case holds here. The sum of the two diagrams in figure \ref{fig:match_one_loop_sca_1} indeed vanishes. Then, the same diagrams with a kinetic or an external energy insertion vanish individually, due to an odd number of the static propagators involved. The last set (iv) equally vanishes, since they are an iteration of the one-loop diagrams (i), as shown in \cite{Peter:1997me}. 

\begin{figure}[t!]
    \centering
    \includegraphics[scale=0.55]{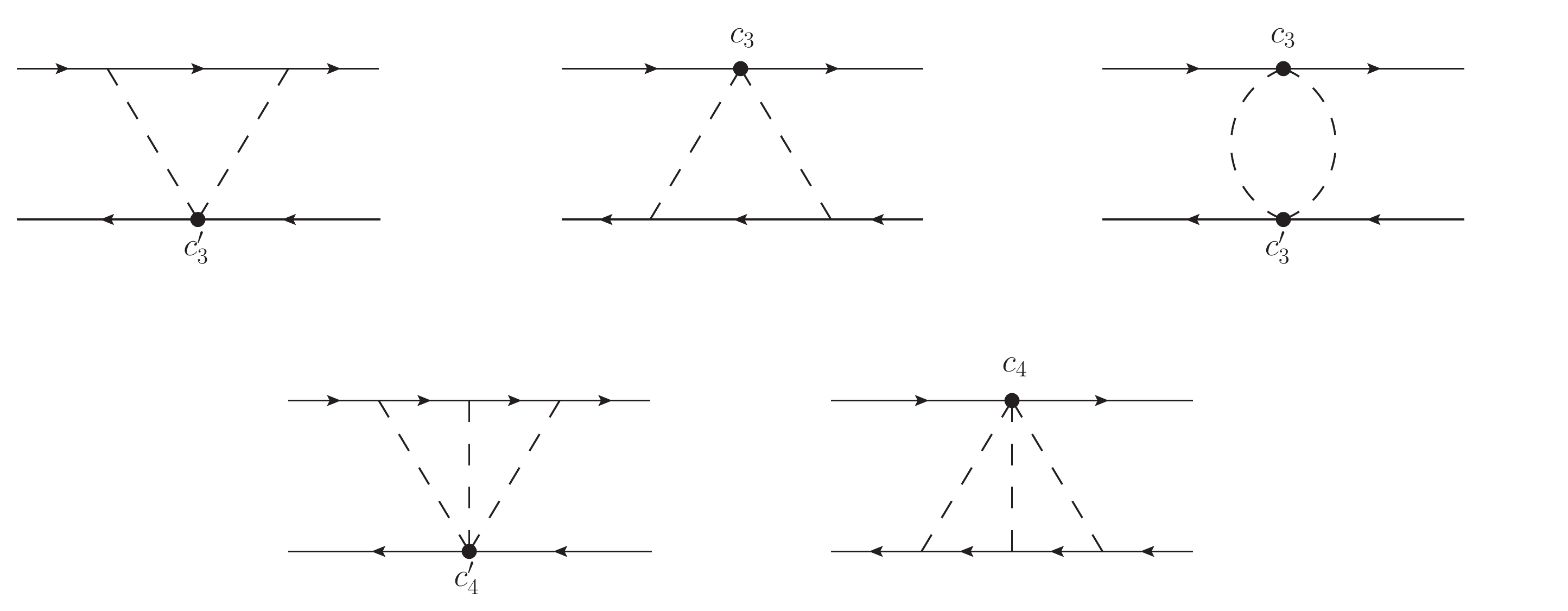}
    \caption{Upper row: one-loop diagrams with two-scalar vertices. The sum of the first and second diagrams vanishes, the third diagram is of order $M \alpha^5$. Lower row: examples of two-loop diagrams involving one three-scalar-vertex, they contribute beyond the desired accuracy $M \alpha^4$.}
    \label{fig:match_one_loop_sca_2bis}
\end{figure}

In addition to the previous class of diagrams, we have to consider possible contributions arising from other topologies, namely those that are induced by the interactions between fermions (antifermions) and two or three scalars. Before discussing the diagrams in some detail, let us remind that the coefficients $c_3(c'_3)$ and $c_4(c'_4)$ vanish at tree-level, so that $c_3,c_4=\mathcal{O}(\alpha),\mathcal{O}(\lambda)$ at least. Actually, as far as $c_3(c'_3)$ is concerned, one finds that the matching coefficients go like $\mathcal{O}(\alpha^2)$, because the tree-level topology is reproduced in the NRY, and then there is no contribution to $c_3(c'_3)$ at order $\alpha$.  The one-loop diagrams involving the vertices with two scalar fields are collected in figure \ref{fig:match_one_loop_sca_2bis} (upper row). 
They all contribute at order $M \alpha^5$ or higher. 
Finally, two example diagrams with a three-scalar-fields vertex are given in figure \ref{fig:match_one_loop_sca_2bis} (lower row). By applying the power counting one sees that they all go beyond the accuracy of this work, $M \alpha^{9/2}$. All other diagrams involving $c_4$ and $c_3$ are further suppressed. 

\subsection{Master integrals}

Here we provide explicit analytic results for some of the 1-loop integrals that we encountered in the course of calculations done
in this work
\begin{align}
	 \int \frac{d^d l}{(2\pi)^d} \frac{1}{(-l^0 + i \eta)(l^2- m^2 + i \eta)} &= - i  4^{\varepsilon-2} \pi^{\varepsilon - 3/2} \Gamma \left ( \varepsilon - \frac{1}{2} \right ) m^{1 - 2\varepsilon}, \\
 	 \int \frac{d^d l}{(2\pi)^d} \frac{1}{(-l^0 + \Delta E + i \eta)(l^2 + i \eta)} & = \frac{i \pi^{\varepsilon -2}}{4} \Gamma(1 - \varepsilon) \Gamma(2 \varepsilon - 1)  \left ( - \Delta E - i \eta \right )^{1-2 \varepsilon}.
\end{align}

\newpage
\bibliographystyle{hieeetr}
\bibliography{paper.bib}
\end{document}